\documentclass[journal]{IEEEtran}
\usepackage[english]{babel}
\usepackage[T1]{fontenc}
\usepackage[utf8]{inputenc}

\usepackage{amsmath,amssymb,amsfonts}
\usepackage{xcolor}
\usepackage{colortbl}
\usepackage{booktabs}
\usepackage{graphicx}
\usepackage{tabularx}
\usepackage{multirow}

\usepackage{placeins}
\usepackage{float}

\usepackage{pgfplots}
\pgfplotsset{compat=newest}
\pgfplotsset{plot coordinates/math parser=false}
\newlength\figureheight
\newlength\figurewidth 

\graphicspath{{images/}}

\definecolor{pastellGreen}{rgb}{0.40,0.80,0.20}%
\definecolor{pastellBlue}{rgb}{0.60,0.80,1.00}%
\definecolor{pastellOrange}{rgb}{1,0.80,0.4}%
\definecolor{pastellRed}{rgb}{1.00,0.47,0.30}%
\definecolor{pastellViolet}{rgb}{0.6,0.6,0.80}%

\ifCLASSINFOpdf
\else
\fi
\begin{document}
%
\title{Dynamic Non-Regular Sampling Sensor Using Frequency Selective Reconstruction}
%
%
%

\author{Markus~Jonscher, Jürgen~Seiler,~\IEEEmembership{Senior Member, IEEE}, Daniela Lanz,~\IEEEmembership{Student Member, IEEE}, Michael Schöberl, Michel Bätz,~\IEEEmembership{Student Member, IEEE}, and~André~Kaup,~\IEEEmembership{Fellow, IEEE}
\thanks{M. Jonscher, J. Seiler, D. Lanz, M. Bätz, and A. Kaup are with the Chair of Multimedia Communications and Signal Processing, University of Erlangen-Nürnberg (FAU), Cauerstr. 7, 91058 Erlangen, Germany (e-mail: markus.jonscher@fau.de; juergen.seiler@fau.de; daniela.lanz@fau.de; michel.baetz@fau.de; andre.kaup@fau.de). M. Schöberl is with the Fraunhofer Institute for Integrated Circuits, Am Wolfsmantel 33, 91058 Erlangen, Germany (e-mail: michael.schoeberl@iis.fraunhofer.de)

Copyright \copyright\ 2018 IEEE. Personal use of this material is permitted. However, permission to use this material for any other purposes must be obtained from the IEEE by sending an email to pubs-permissions@ieee.org.
}}

%
%

\markboth{}%
{Shell \MakeLowercase{\textit{et al.}}: Bare Demo of IEEEtran.cls for Journals}
%



\maketitle

\begin{abstract}
Both a high spatial and a high temporal resolution of images and videos are desirable in many applications such as entertainment systems, monitoring manufacturing processes, or video surveillance. Due to the limited throughput of pixels per second, however, there is always a trade-off between acquiring sequences with a high spatial resolution at a low temporal resolution or vice versa.  In this paper, a modified sensor concept is proposed which is able to acquire both a high spatial and a high temporal resolution. This is achieved by dynamically reading out only a subset of pixels in a non-regular order to obtain a high temporal resolution. A full high spatial resolution is then obtained by performing a subsequent three-dimensional reconstruction of the partially acquired frames. The main benefit of the proposed dynamic readout is that for each frame, different sampling points are available which is advantageous since this information can significantly enhance the reconstruction quality of the proposed reconstruction algorithm. Using the proposed dynamic readout strategy, gains in PSNR of up to $\mathbf{8.55}$~dB are achieved compared to a static readout strategy. Compared to other state-of-the-art techniques like frame rate up-conversion or super-resolution which are also able to reconstruct sequences with both a high spatial and a high temporal resolution, average gains in PSNR of up to $\mathbf{6.58}$~dB are possible.
\end{abstract}


%
\IEEEpeerreviewmaketitle

\section{Introduction}
\label{sec:introduction}
Typical complementary metal-oxide-semiconductor camera sensors have many identical pixels arranged in a fixed regular array. These elements are evenly spaced and define the sensor sampling pattern and therefore the maximum spatial resolution of the camera. The number of pixels (per second) also defines the required throughput of processing, compression, and storage. The complexity of the camera electronics, weight, and cost is therefore closely coupled to the number of pixels that needs to be processed. For many applications such as video surveillance or entertainment systems, an ongoing pursuit for both a high spatial and a high temporal resolution can be discovered. A straightforward solution to enhance the spatial resolution is to increase the number of pixels. However, besides all the limitations mentioned before, increasing the number of pixels and therefore reducing the size of the individual pixels is only possible up to a certain point due to photometric limits~\cite{Brueckner2013}. The temporal resolution is restricted by the number of pixels that can be processed per second. Hence, conventional video cameras capture sequences either with a high spatial resolution at a low frame rate or vice versa.

In order to overcome this trade-off, the concept of a non-regular sampling sensor has been proposed in~\cite{Schoeberl2011} where a low resolution sensor with a regular sampling pattern can be used to obtain sequences with a high temporal resolution, however, at a low spatial resolution. In order to obtain a high spatial resolution, this low resolution sensor is shielded in a way that three quarters of every large pixel are non-regularly masked. In doing so, a sequence is acquired with twice the resolution in both spatial dimensions. However, three quarters of all pixels of these high resolution frames are missing and have to be reconstructed by a subsequent reconstruction algorithm. Unfortunately, this non-regular sampling pattern has the major disadvantage that it is static and cannot be changed dynamically over time in order to sample all positions of a scene. 

\begin{figure}
	\centering
	\def\svgwidth{\columnwidth}
	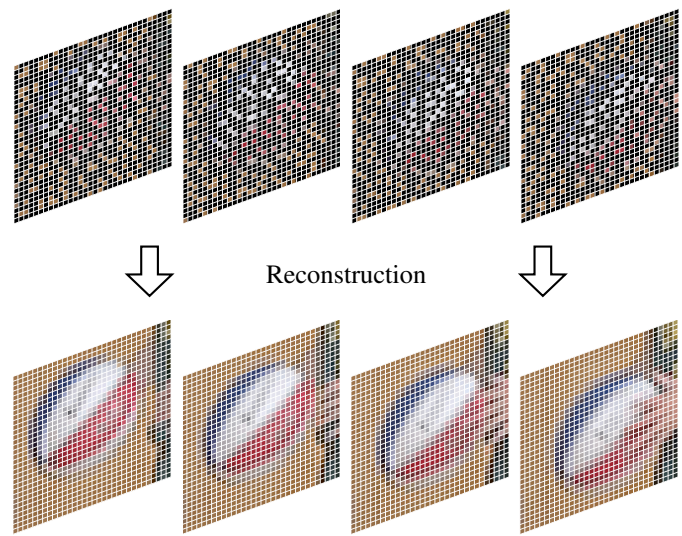
	\caption{Concept of the proposed dynamic non-regular sampling sensor: Less pixels are read out in a non-regular order to obtain a high temporal resolution. A subsequent reconstruction of the missing pixels yields the full high spatial resolution. Using the proposed dynamic pixel readout, it is ensured that each position of a scene can be sampled.}
	\label{fig:motivation-dynamic-sampling}
\end{figure}
Therefore, a novel dynamic non-regular pixel readout strategy is proposed in this paper. The concept is shown in Figure~\ref{fig:motivation-dynamic-sampling}. It allows for the construction of an image sensor that is able to acquire video sequences with both a high spatial and a high temporal resolution. This is achieved by reading out less pixels per frame in a non-regular order which leads to a high temporal resolution, since more frames per second can be acquired. The full high spatial resolution is then obtained by a subsequent reconstruction of the missing pixels in the frames where only a subset of pixels is available. For the reconstruction, a three-dimensional algorithm is proposed that is able to reconstruct non-regularly sampled video data with a high quality. 
In previous work such as~\cite{Schoeberl2011, Jonscher2014b, Jonscher2014a}, a non-regular sampling sensor has been employed that utilizes a physical mask in order to obtain the non-regular sampling pattern. The proposed non-regular sampling pattern in this paper, however, is obtained by a specific readout technique and a physical masking is therefore not required. This way, major disadvantages such as diffraction that a physical mask would cause can be avoided. Another advantage of using only the specific readout strategy is that the sampling pattern can be easily varied over time, so that different sampling points can be used for enhancing the reconstruction quality.
By using the proposed dynamic sampling concept in combination with the proposed reconstruction algorithm, a video sequence with both a high spatial and a high temporal resolution can be acquired. Additionally, other methods from frame rate up-conversion (FRUC)~\cite{Haan2010,Baetz2017} and super-resolution (SR)~\cite{Park2003,Karimi2014,Nasrollahi2014,Walha2016,Lim2017,Dong2013} which are also able to reconstruct such sequences are used for comparison.

This article starts with a review of several applications for non-regular sampling in general and the non-regular sampling sensor with a static sampling pattern in particular. Section~\ref{sec:dynamic-sampling} then introduces the proposed dynamic non-regular sampling sensor concept followed by the proposed three-dimensional reconstruction algorithm for non-regularly sampled video data in Section~\ref{sec:3D-FSR}. Simulation results and visual examples are given in Section~\ref{sec:results} and Section~\ref{sec:conclusion} concludes this paper.

\section{Non-Regular Sampling}
\label{sec:non-reg-sampling}

Since conventional sensors perform a regular two-dimensional sampling, the resolution is limited by aliasing. However, it has been shown in~\cite{Hennenfent2007} and~\cite{Maeda2009} that the visual influence of aliasing can be reduced by a non-regular sampling. There exist many systems where a non-regular sampling occurs either by default due to the system's characteristics or on purpose in order to exploit certain properties of the non-regularity. 
One reason why a non-regular sampling is employed instead of a regular sampling, is the advantage that the spectrum of such a non-regular sampling pattern consists of a dominant peak at the direct current component while all other frequencies contribute as a noise-like floor. Since image signals can be sparsely represented in the Fourier domain which means that most of the signal energy is concentrated into only a small number of transform coefficients while all other coefficients are zero or close to zero, the dominant basis functions can still be identified after a non-regular sampling. The consequences and advantages from a non-regular sampling compared to a regular sampling can also be found in more detail in~\cite{Seiler2015}. 
The different non-regular sampling applications can be grouped into two categories. The first one contains all sampling systems that use a static sampling. That is to say that the sampling points do not change over time. The second group gathers all systems that are able to vary the sampling points dynamically over time.

\subsection{Static Sampling}
The most simple static non-regular sampling pattern to think of is caused by a regular high resolution sensor where some of the pixels are damaged. Such a defect may be caused by manufacturing, aging, or radiation. Imaging sensors with few defects are quite common and instead of discarding the chip, an interpolation is used to fill the missing pixel values. This reconstruction task can directly be described as a non-regular sampling problem. With an increasing number of defects, however, this interpolation becomes very challenging as damaged pixels tend to group in clusters and has been evaluated in detail in~\cite{Schoeberl2011a}. 
A similar problem is considered in~\cite{Sen2009} where a reconstruction algorithm based on compressed sensing is used for image sensors with many defects. The readout of this sensor requires just as much time and power as a fully functional device, since all transistors, multiplexers and A/D converters are driven with full power and only in a digital stage the defects can be sorted out.
Another system for a static non-regular sampling is shown in~\cite{Slagle2009} where a scrambled bundle of optical fibers of different diameters is used to non-regularly map the light path between optics and sensor. At the display, a second corresponding random face plate is used for inverse mapping.

\begin{figure*}
	\centering
	\def\svgwidth{\textwidth}
	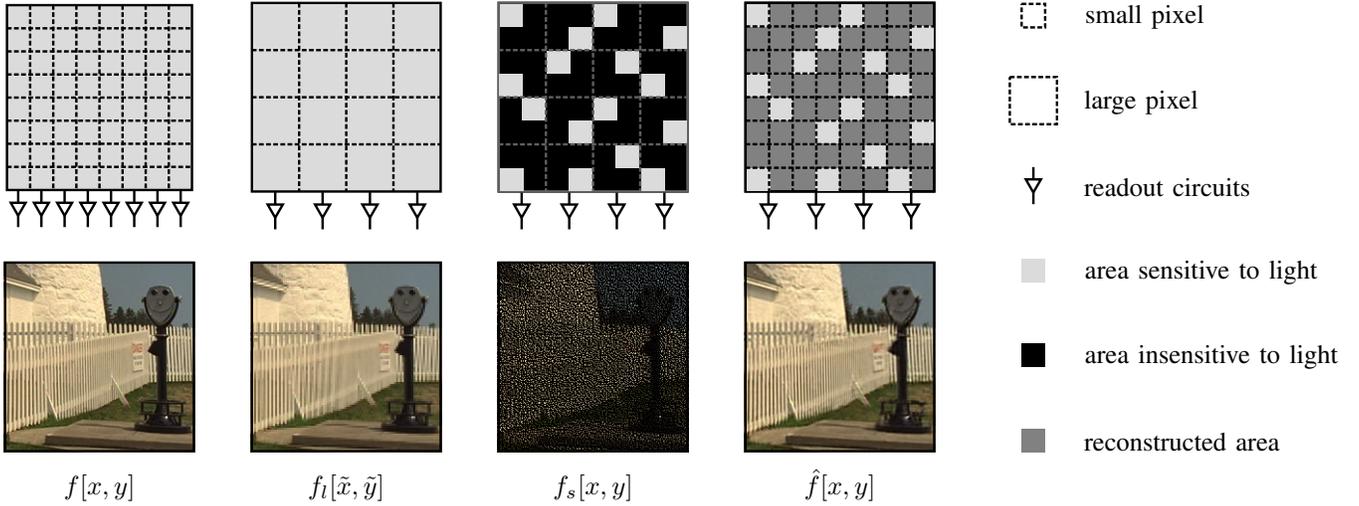
	\caption{Principle of the non-regular sampling sensor. From left to right: High resolution image sensor yielding the high resolution image $f[x,y]$. Low resolution image sensor giving the low resolution image $f_l[\tilde{x},\tilde{y}]$. Low resolution image sensor covered by a static non-regular sampling pattern leading to the sampled image $f_s[x,y]$. High resolution image $\hat{f}[x,y]$ obtained after the reconstruction of the missing pixels in $f_s[x,y]$.}
	\label{fig:non-reg-principle}
\end{figure*}
In~\cite{Schoeberl2011}, a modified image sensor concept has been proposed that allows for the reconstruction of an image with a higher spatial resolution than the original sensor is able to acquire. The basic idea of this approach is shown in Figure~\ref{fig:non-reg-principle}. On the left, a high resolution sensor that yields the high resolution image $f[x,y]$ is illustrated. The area of the sensor that is sensitive to light is denoted in light-gray and $(x,y)$ depict the spatial coordinates on the high resolution grid. This sensor would be the best option if there were no restrictions in resources. To reduce  hardware costs, energy consumption, or storage space of the data acquisition, however, an image sensor with less pixels and therefore fewer readout circuits may be necessary to employ. The resulting image captured by such a low resolution sensor leads to the image $f_l[\tilde{x},\tilde{y}]$ of lower quality, where $(\tilde{x},\tilde{y})$ depict the spatial coordinates on the low resolution grid. It can be seen that the regular placement of large pixels leads to the well-known effect of aliasing where high frequencies get mapped to low frequencies which results in strong artifacts. Nevertheless, it is desirable to retain all of the advantages of a low resolution imaging sensor and still obtain a high resolution output. Therefore, this low resolution sensor gets covered with a non-regular sampling pattern which ensures that each large pixel is divided into four quadrants, where three of them are non-regularly covered. As a consequence, only one quarter of the large pixel is sensitive to light anymore which leads to a high resolution image $f_s[x,y]$, where due to the masking, only a subset of pixels is available on the high resolution grid. A subsequent reconstruction algorithm is then needed for the reconstruction of these missing pixels which finally leads to the reconstructed high resolution image $\hat{f}[x,y]$. The optimal layout of such a non-regular sampling pattern is still an ongoing research topic, however, it has been shown in~\cite{Jonscher2014b} that it is sufficient to employ sampling patterns that are non-regular only on a small scale and are then repeated to match any image sensor size that is required. This will also simplify the manufacturing process.
Up to now, the reconstruction of the partially available image data that is captured by a non-regular sampling sensor has been performed by the two-dimensional frequency selective extrapolation (2D-FSE)~\cite{Seiler2010} or the enhanced two-dimensional frequency selective reconstruction (2D-FSR) in~\cite{Seiler2015}. Recently, it has been shown in~\cite{Jonscher2014a,Jonscher2015} that the reconstruction quality can be enhanced by exploiting spatial or temporal correlations between neighboring views or frames. A further enhancement of the reconstruction quality can be expected by the use of algorithms that directly reconstruct volumes~\cite{Meisinger2007a,Seiler2016b,Garcia2010}. The non-regular sampling sensor, however, has the major disadvantage that the sampling pattern cannot be changed dynamically over time.

\subsection{Dynamic sampling}
\begin{figure}[t]
	\def\svgwidth{\columnwidth}
	\hspace{-0.4cm}
	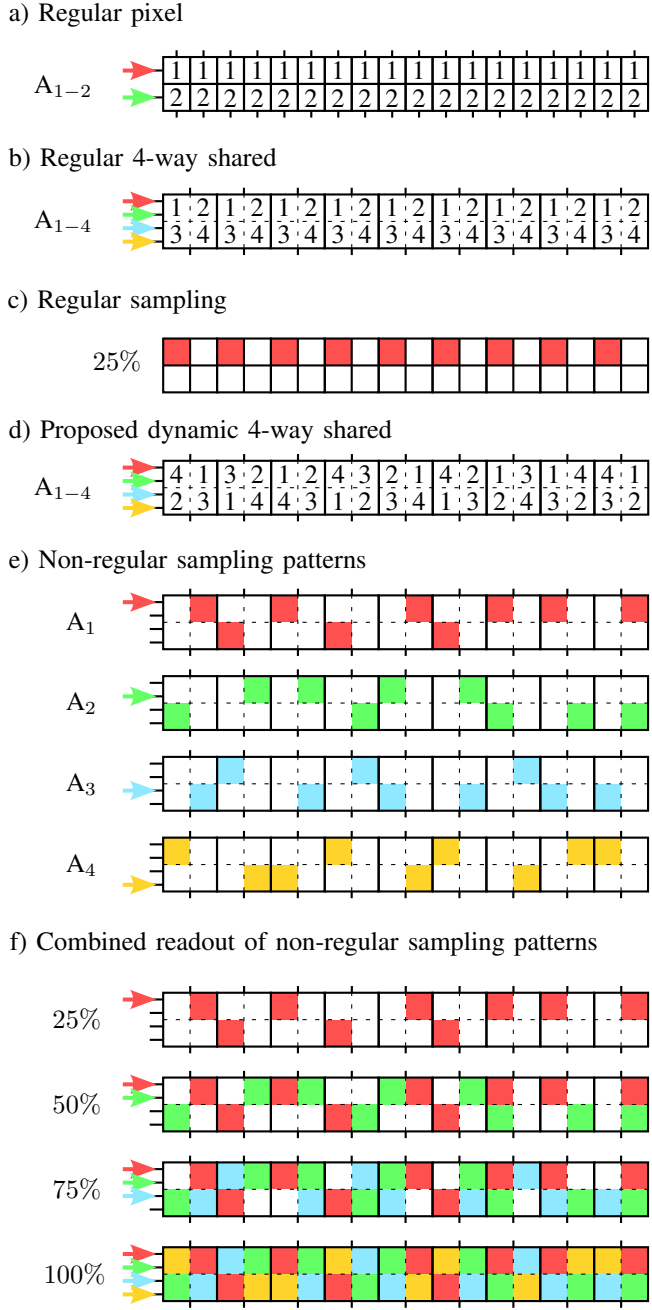
	\caption{Different readout strategies and sampling patterns: a) Regular pixels and address lines, b) 4-way shared regular pixel architecture~\cite{Mori2004}, c) Regular sampling pattern of a) and b), d) Proposed 4-way shared non-regular pixel architecture, e) Exemplary sampling patterns of the proposed 4-way shared non-regular pixels, f) Non-regular readout patterns for readout of $25\%$, $50\%$, $75\%$, and full sensor readout of the proposed architecture.}
	\label{fig:dynamic-sampling-pattern}
\end{figure}

The adjustment of the pixel readout at run time is commonly used in high resolution still image cameras that also feature video modes with reduced resolution. Since the processing throughput in cameras is limited to a certain number of pixels per seconds, only a subset of pixels is read out and processed at typical video frame rates. 
As shown in Figure~\ref{fig:dynamic-sampling-pattern}a, when using a regular pixel architecture, there is a single address line for each row of pixels. The activation of address line 1 activates all transfer gates in the pixels of row 1 which allows for the following two options of sampling. By applying cropping~\cite{Potter2003}, only a certain region of the sensor is used and other address lines are simply not activated. This leads, however, to a very narrow field-of-view of the camera and is therefore only rarely used. The second sampling option is line skipping, where only every other row is read out. Additionally, by deactivating every other readout circuit, this effectively performs a horizontal and vertical sampling where a typical pattern with a sampling density of $25\%$ can be seen in Figure~\ref{fig:dynamic-sampling-pattern}c. Depending on the readout, different sampling patterns can be realized.

Another system that employs dynamic sampling is presented in \cite{Bub2010} where a Digital Micromirror Device is utilized in order to control the light that reaches the sensor. The exposure time of the camera is subdivided into slices and in each slice only some of the pixels are activated. This proposed setup can also be used to perform a pseudo non-regular sampling. The optical setup with Digital Micromirror Device devices, however, creates a costly and large camera system.

A further approach for achieving dynamic non-regular sampling patterns is shown in~\cite{Lin2012}. This system has all samples available in memory and only random pixels are used for further processing. For the scenario regarded in this paper, capturing, discarding, and reconstructing pixel values is not an option and the direct use of the captured samples is to be preferred.

\section{Proposed Dynamic Non-Regular Sampling Sensor Concept}
\label{sec:dynamic-sampling}

In this paper, a novel approach for the construction of image sensors that allow for a dynamic non-regular sampling is proposed. It builds upon the 4-way shared pixel sensor presented in~\cite{Mori2004}. The order for each group of four pixels, however, is randomized. The basic idea of the regular 4-way shared pixel sensor is that multiple pixels share some of the required electronics. The area of a traditional CMOS pixel cell can be classified into a photosensitive area and a non-sensitive area which contains the electronics for reading the pixels. The ratio of the sensitive area to the total pixel area is called fill factor. A straightforward shrinking of the pixel size in order to achieve a higher resolution results in a poor fill factor and therefore a bad sensitivity of the camera. Instead of $4$ transistors per pixel, the reported pixel cell in~\cite{Mori2004} only requires $1.75$ transistors per pixel and thus a reasonably high fill factor can be achieved. The resulting group of four pixels now requires four address lines and feeds to a single column bus for readout as can be seen in Figure~\ref{fig:dynamic-sampling-pattern}b. A typical regular sampling pattern with a density of $25\%$ of such a sensor when only the first address line is activated is shown in Figure~\ref{fig:dynamic-sampling-pattern}c.

For the construction of dynamic non-regular sampling patterns, the order of the pixels for each group of four pixels of such a regular 4-way shared pixel sensor is randomized as can be seen in Figure~\ref{fig:dynamic-sampling-pattern}d. In doing so, only a different wiring is necessary. This way, a non-regular sampling pattern with a sampling density of $25\%$ can be achieved where with the activation of address line $\mathrm{A}_1$ all the pixels marked with $1$ are read out. The four resulting sampling patterns $\mathrm{A}_1-\mathrm{A}_4$ of two rows of such a sensor are depicted in Figure~\ref{fig:dynamic-sampling-pattern}e, where for each of the patterns only a single address line is activated. On the whole sensor a sampling density of $25\%$ can be performed using a large number of non-regular patterns, since each group of two rows is independent and can select from one of four possible patterns. In doing so, non-regularity can be ensured in both spatial and temporal direction.

It is also possible to achieve a sampling density of $50\%$ or $75\%$ by reading out multiple address lines. This gives again a large number of combinations for the readout of the proposed image sensor. The full readout with $100\%$ is just as expensive as the full readout of a regular 4-way shared pixel design~\cite{Mori2004} from Figure~\ref{fig:dynamic-sampling-pattern}b. After the readout and A/D conversion, however, the pixels need to be put in correct order. As the non-regularity is only limited to a $2\times2$ area, the rearrangement of pixel values can be carried out locally and with little memory.

Compared to a static non-regular sampling where for each frame of a sequence the missing pixels are at the same position, using the proposed dynamic non-regular sampling leads to a varying sampling of the pixels over time. As can be seen in Figure~\ref{fig:dynamic-sampling-pattern}e for example, for a sampling density of $25\%$, complementary patterns can be read out. This is achieved by reading out first $\mathrm{A}_1$, then $\mathrm{A}_2$, and so on. In this case, it is ensured that every position gets sampled at least once within $4$ frames. In general, that is to say that especially for areas with no or little motion, a static sampling cannot cover all the information of the scene. Using the proposed dynamic sampling, however, the information of almost the whole scene can be acquired, but spread over several frames. This way, sampling points with distinct information can support a subsequent reconstruction method.

In general, there are many different non-regular sampling patterns possible.
\begin{figure}
	\centering
	\def\svgwidth{\columnwidth}
	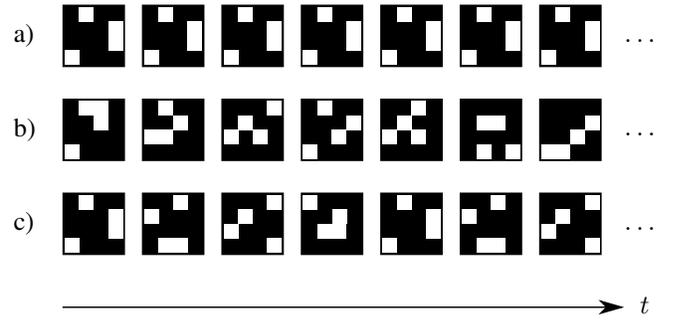
	\caption{a) Static sampling. b) Completely non-regular sampling in all three dimensions. c) Proposed dynamic non-regular sampling.}
	\label{fig:different-sampling-patterns}
\end{figure}
\begin{figure*}
	\def\svgwidth{\textwidth}
	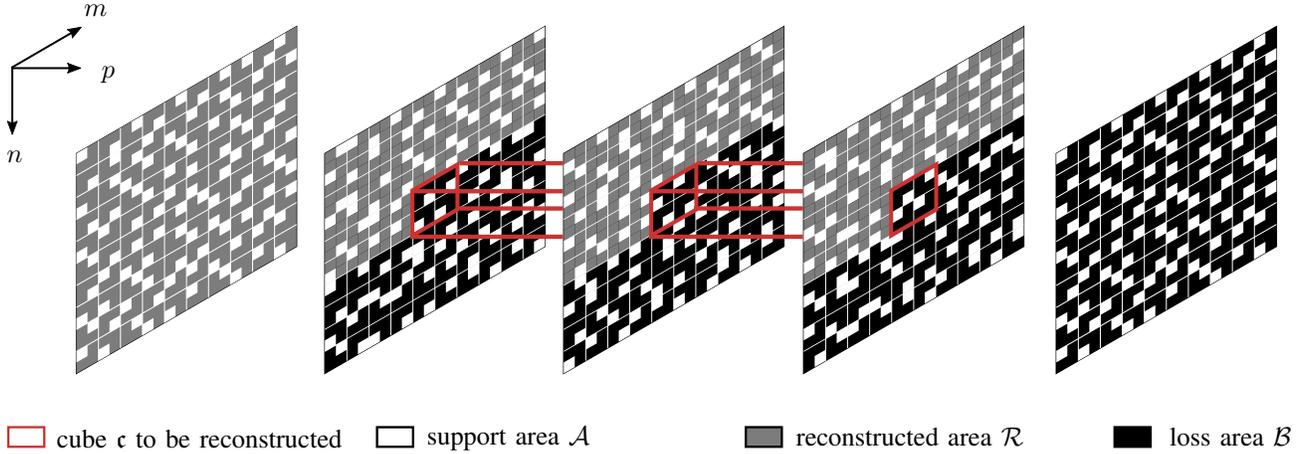
	\caption{Reconstruction area $\mathcal{L} = \mathcal{A} \cup \mathcal{B} \cup \mathcal{R}$ consisting of the support area $\mathcal{A}$, the loss area $\mathcal{B}$, and the area $\mathcal{R}$ which contains all the pixels that have been already reconstructed. The cube $\mathfrak{c}$ is the cube actually to be reconstructed and marked in red in the center of the reconstruction area $\mathcal{L}$.}
	\label{fig:extrapolation-area}
\end{figure*}
In Figure~\ref{fig:different-sampling-patterns}, different sampling patterns are illustrated for a sampling density of $25\%$. A static sampling with no variation in temporal direction can be seen in Figure~\ref{fig:different-sampling-patterns}a. A straightforward way for a dynamic sampling is shown in Figure~\ref{fig:different-sampling-patterns}b where the sampling is completely non-regular in all three dimensions. As the proposed realization of a dynamic readout is restricted due to the wiring, a non-regular sampling as can be seen in Figure~\ref{fig:different-sampling-patterns}c is considered in this paper. In this case, the sampling is not completely non-regular, since for each group of four pixels one pixel is taken. However, as natural video sequences are non-stationary, this is advantageous, since this kind of somewhat evenly distributed sampling is better for adjusting to changes in statistical properties of such sequences.

\section{Three-Dimensional Frequency Selective Reconstruction}
\label{sec:3D-FSR}
For an effective reconstruction of erroneous video data, the three-dimensional frequency selective extrapolation (3D-FSE) has been proposed in~\cite{Meisinger2007a}. 3D-FSE, however, has been originally designed to reconstruct block losses or connected block losses and it is not very well suited for the reconstruction of non-regularly sampled video data. Therefore, the three-dimensional frequency selective reconstruction (3D-FSR) is proposed in this paper.

In the scenario considered in this paper, a video sequence $v[x,y,t]$ gets captured by a non-regular sampling sensor where $(x,y)$ depict the spatial coordinates and $t$ the temporal coordinate. Depending on the sampling density, $v[x,y,t]$ contains a different number of missing pixels that have to be reconstructed in order to obtain the full high resolution sequence. 3D-FSR performs a block-wise reconstruction and therefore splits $v[x,y,t]$ into equally-sized cubes~$\mathfrak{c}$. Already reconstructed neighboring cubes can be reused in order to utilize the reconstructed pixels in the reconstruction process for other cubes. In total, all of the cubes~$\mathfrak{c}$ that contain missing pixels have to be processed. 
In general, there exist many different orders for processing these cubes. A straightforward way is to process the cubes in a line scan order~\cite{Meisinger2007a}. This line scan order, however, is not suitable to exploit all available information from all directions effectively which may lead to annoying error propagation artifacts.
In order to cope with this drawback, an optimized processing order has been presented in~\cite{Seiler2016b}. This processing order has been originally designed for the reconstruction of large arbitrarily shaped areas of missing pixels and a parallel processing of the cubes~$\mathfrak{c}$. It is, however, not very well suited for the scenario considered in this paper, since due to the distribution of the sampling points no large holes can occur and the processing order therefore gets reduced to almost a line scan order. 
In~\cite{Seiler2015}, another processing order has been proposed for 2D-FSR which takes the local density of available pixels into account and is therefore better suited for the reconstruction of non-regularly sampled image data.

For the reconstruction of non-regularly sampled video data, a similar processing order as in~\cite{Seiler2015} may be advantageous, since cubes with many known pixels should be processed first in order to support the reconstruction of cubes with fewer pixels. In doing so, the sampling pattern gets first lowpass filtered with a three-dimensional Gaussian window.
The pixel values within each cube are summed up and the cubes are then processed in decreasing order of the result of the summation. After determining the processing order of the individual cubes, the actual reconstruction starts with the cube which has the most available pixels.

As shown in Figure~\ref{fig:extrapolation-area}, not only the cube $\mathfrak{c}$ is regarded for the reconstruction but also a certain neighborhood. This whole area is called the reconstruction area $\mathcal{L}$ with the coordinates $(m,n,p)$ and it is of size $M\times N\times P$. The reconstruction area $\mathcal{L}$ is divided into several areas where all originally taken pixels are gathered in the support area $\mathcal{A}$. All pixels that are missing and have to be reconstructed are subsumed in the loss area $\mathcal{B}$ and all pixels from neighboring cubes that have been reconstructed before belong to the area $\mathcal{R}$.

The objective of 3D-FSR now is to generate the complex-valued signal model
\begin{equation}
	g[m,n,p] = \sum\limits_{(k,l,q)\in\mathcal{K}}\hat{c}_{(k,l,q)} \cdot \varphi_{(k,l,q)}[m,n,p].
\end{equation}
for $s[m,n,p]$, which corresponds to the signal that is acquired in area $\mathcal{L}$, as an iterative superposition of Fourier basis functions
\begin{equation}
	\varphi_{(k,l,q)}[m,n,p] = \mathrm{e}^{\mathrm{j}\frac{2\pi}{M}km}\mathrm{e}^{\mathrm{j}\frac{2\pi}{N}ln}\mathrm{e}^{\mathrm{j}\frac{2\pi}{P}qp}
\end{equation}
weighted by the expansion coefficients $\hat{c}_{(k,l,q)}$. The set $\mathcal{K}$ contains all the utilized basis functions. That is to say that in each iteration $\nu$, one basis function gets selected which is then added to the model with the appropriate weight. The approximation residual
\begin{equation}
	r^{(\nu-1)}[m,n,p] = s[m,n,p] - g^{(\nu-1)}[m,n,p]
\end{equation}
with respect to the previous iteration is then computed at the beginning of each iteration $\nu$. Initially, the model $g^{(0)}[m,n,p]$ is set to zero. Afterwards, the projection coefficients
\begin{equation}
	p^{(\nu)}_{(k,l,q)} = \frac{\sum\limits_{(m,n,p)\in\mathcal{L}}\hspace*{-0.3cm}r^{(\nu-1)}[m,n,p] \cdot \varphi^{\ast}_{(k,l,q)}[m,n,p] \cdot w[m,n,p]} {\sum\limits_{(m,n,p)\in\mathcal{L}}\left|\varphi_{(k,l,q)}[m,n,p]\right|^2 \cdot w[m,n,p]}
\end{equation}
are calculated which result from a weighted projection of the residual onto all basis functions. Therefore, the weighting function
\begin{equation}
	w[m,n,p] = \begin{cases}
		\rho[m,n,p] & \text{for}\ (m,n,p)\in\mathcal{A} \\
		\delta\rho[m,n,p] & \text{for}\ (m,n,p)\in\mathcal{R} \\
		0 & \text{for}\ (m,n,p)\in\mathcal{B}
	\end{cases}
\end{equation}
with the three-dimensional exponentially decreasing function
\begin{equation}
	\rho[m,n,p] = \hat{\rho}^{\sqrt{\left(m-\frac{M-1}{2}\right)^2 + \left(n-\frac{N-1}{2}\right)^2 + \left(p-\frac{P-1}{2}\right)^2}}
\end{equation}
is applied in order to control the influence of each pixel on the model generation depending on its position. That is to say that pixels farther away from the center of the cube $\mathfrak{c}$ get a lower weight than pixels in the direct neighborhood. The parameter~$\hat{\rho}$ controls the speed of this decay. Another important property of the weighting function $w[m,n,p]$ is that all pixels from the loss area $\mathcal{B}$ are excluded from the model generation. Since pixels from neighboring cubes that have been reconstructed before are not as reliable as originally acquired pixels, they are weighted by an additional reduction factor $\delta$ in the range between zero and one. 

The basis function that gets selected in iteration $\nu$ and added to the model is the one that minimizes the distance between the residual and the corresponding projection onto the basis function weighted by $w[m,n,p]$. The index of this basis function is calculated by
\begin{equation}
\begin{split}
	(u,v,z) = \underset{(k,l,q)}{\mathrm{argmax}}\Bigg(\left|p^{(\nu)}_{(k,l,q)}\right|^2 \cdot w_f^{(\Omega)}[k,l,q] \cdot \\ \sum\limits_{(m,n,p)\in\mathcal{L}}\left|\varphi_{(k,l,q)}[m,n,p]\right|^2w[m,n,p]\Bigg).
\end{split}
\end{equation}
The additional frequency prior
\begin{equation}
	w_f^{(\Omega)}[k,l,q] = \left(1 - \sqrt{2} \cdot \sqrt{\frac{\tilde{k}^2}{M^2} + \frac{\tilde{l}^2}{N^2} + \frac{\tilde{q}^2}{P^2}}\right)^{2\alpha(\Omega)}
\end{equation}
is utilized for the selection process in order to favor certain basis functions depending on the number of available pixels in the reconstruction area $\mathcal{L}$. For example, if $\mathcal{L}$ only contains very few originally taken pixels, a low-frequency solution should be preferred in order to avoid artifacts. If, however, many pixels are available, the suppression of high-frequency basis functions should be small. Using the measure
\begin{equation}
	\Omega = \frac{\sum\limits_{(m,n,p)\in\mathcal{A}\cup\mathcal{R}}w[m,n,p]}{\sum\limits_{(m,n,p)\in\mathcal{L}}\rho[m,n,p]}
\end{equation}
of the effective data, the frequency prior $w_f^{(\Omega)}[k,l,q]$ can be made adaptive to the number of available pixels. The weighting function $w[m,n,p]$ is again used to control the influence of each pixel depending on the distance to the cube to be reconstructed. If $\hat{\rho}=1$, $\Omega$ would just be the number of available pixels in the reconstruction area $\mathcal{L}$. The mapping function
\begin{equation}
	\alpha(\Omega) = -\frac{\log(\Omega)}{\tau}
\end{equation}
is used in the exponent of the frequency prior $w_f^{(\Omega)}[k,l,q]$, since the measure $\Omega$ of the effective data cannot be directly inserted. The parameter $\tau$ is introduced in order to control the adaptive suppression of high-frequency basis functions depending on the number of available pixels in the reconstruction area $\mathcal{L}$.

The expansion coefficient
\begin{equation}
	\hat{c}^{(\nu)}_{(u,v,z)} = \gamma \cdot p^{(\nu)}_{(u,v,z)}
\end{equation}
of the selected basis function that gets added to the signal model $g[m,n,p]$ is calculated using the orthogonality deficiency compensation factor $\gamma$ as proposed in~\cite{Seiler2008}, since the projection coefficients $p^{(\nu)}_{(k,l,q)}$ cannot be used directly as an estimate for the influence. After the selection of a suitable basis function and the calculation of the correct weight, the signal model and the residual of the current iteration are updated by
\begin{equation}
	g^{(\nu)}[m,n,p] = g^{(\nu-1)}[m,n,p] + \hat{c}^{(\nu)}_{(u,v,z)} \cdot \varphi_{(u,v,z)}[m,n,p]
\end{equation}
and
\begin{equation}
	r^{(\nu)}[m,n,p] = r^{(\nu-1)}[m,n,p] - \hat{c}^{(\nu)}_{(u,v,z)} \cdot \varphi_{(u,v,z)}[m,n,p].
\end{equation}
After a predefined number of iterations, the real-valued component of the resulting signal model is used to replace all missing pixels in the currently considered cube.

For a faster computation, these steps can also be processed in the frequency domain. In doing so, the basis function selection can be expressed by
\begin{equation}
	(u,v,z) = \underset{(k,l,q)}{\mathrm{argmax}}\left|R^{(\nu-1)}_w[k,l,q]\right|^2
\end{equation}
and the estimation of the corresponding expansion coefficient by
\begin{equation}
	\hat{c}^{(\nu)}_{(u,v,z)} = \gamma \cdot \frac{R^{(\nu-1)}_w[u,v,z]}{W[0,0,0]}.
\end{equation}
$W[k,l,q]$ notes the transformed weighting function $w[m,n,p]$ and $R^{(\nu-1)}_w[k,l,q]$ the transformed weighted residual
\begin{equation}
	r^{(\nu-1)}_w[m,n,p] = r^{(\nu-1)}[m,n,p] \cdot w[m,n,p].
\end{equation}
The update step of the signal model and the residual can also be carried out in the frequency domain by
\begin{equation}
	G^{(\nu)}[u,v,z] = G^{(\nu-1)}[u,v,z] + MNP\hat{c}^{(\nu)}_{(u,v,z)}
\end{equation}
and
\begin{equation}
	R^{(\nu)}_w[k,l,q] = R^{(\nu-1)}_w[k,l,q]-\hat{c}^{(\nu)}_{(u,v,z)} \cdot W[k-u,l-v,q-z].
\end{equation}
Calculating all steps of 3D-FSR in the frequency domain, the computational demand gets significantly reduced and transforms only are required at the beginning and at the end of the model generation.

\section{Simulations \& Results}
\label{sec:results}
In the following, the abilities and properties of the proposed dynamic non-regular sampling concept in combination with the also proposed subsequent reconstruction by 3D-FSR are analyzed. At first, the performance of the proposed 3D-FSR compared to the existing 3D-FSE~\cite{Seiler2016b} is demonstrated. Afterwards, the advantages of the proposed dynamic sampling compared to a static sampling are shown. Additionally, the proposed dynamic readout concept is compared to state-of-the-art frame rate up-conversion (FRUC) and super-resolution (SR) techniques.

\begin{figure}
	\centering
	\def\svgwidth{\columnwidth}
	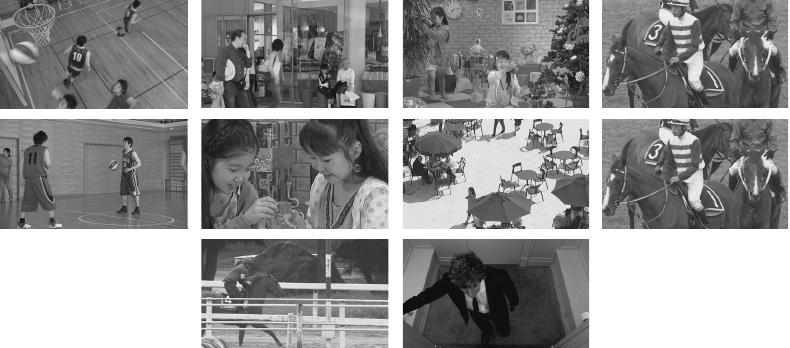
	\caption{Test and training sequences used for simulation. Top: HEVC ClassC test sequences {\itshape BasketballDrill, BQMall, PartyScene, {\normalfont and} RaceHorses}. Center: HEVC ClassD test sequences {\itshape BasketballPass, BlowingBubbles, BQSquare, {\normalfont and} RaceHorses}. Bottom: Training sequences {\itshape Keiba} and {\itshape Mobisode2} from the HEVC data set.}
	\label{fig:test-set}
\end{figure}
The test setup shown in Figure~\ref{fig:test-set} consists of eight sequences from the HEVC data base where for each sequence the first $50$ frames are selected. The first four sequences {\itshape BasketballDrill, BQMall, PartyScene, {\normalfont and} RaceHorses} are from the HEVC ClassC test sequences and are of size $832 \times 480$ pixels. The second four sequences {\itshape BasketballPass, BlowingBubbles, BQSquare, {\normalfont and} RaceHorses}  are from the HEVC ClassD test sequences and are of size $416 \times 240$ pixels. Also shown in Figure~\ref{fig:test-set} are the two sequences  {\itshape Keiba {\normalfont and} Mobisode2} of size $416 \times 240$ also from the HEVC sequences which are used as training sequences. For all simulations and evaluations, only the luminance component is considered.

\begin{figure}
	\centering
	\def\svgwidth{\columnwidth}
	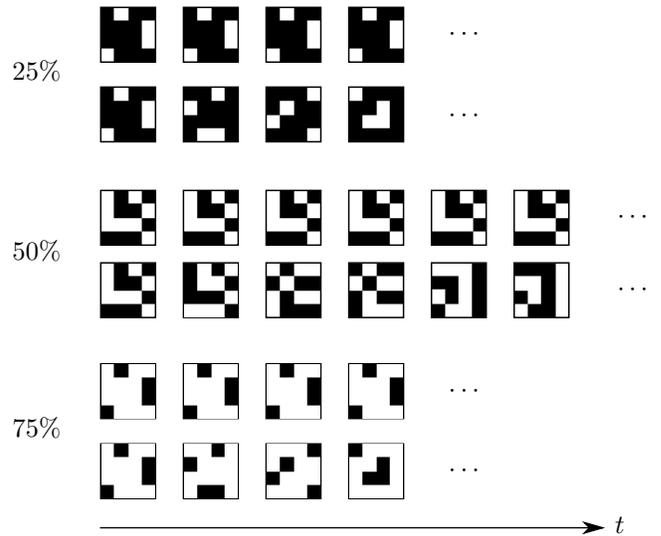
	\caption{Sections of sampling patterns that have been used for simulation. For each sampling density, the upper row shows the static non-regular sampling pattern and the lower row shows the proposed dynamic non-regular sampling pattern. After $4$ or $6$ frames they are repeated in temporal direction.}
	\label{fig:sampling-patterns}
\end{figure}
First, the corresponding readout schemes are applied to the original sequences which have the full high temporal and high spatial resolution. In the case of the proposed non-regular sampling concept this means that various sampling patterns are multiplied with the original sequences in order to simulate a static or dynamic sampling where fewer pixels per frame are read out than the original camera sensor may acquire.
The reconstruction of the missing pixels is evaluated using the two-dimensional frequency selective reconstruction~(2D-FSR) from~\cite{Seiler2015}, a three-dimensional linear interpolation (3D-LIN)~\cite{Watson1994}, the three-dimensional gap filling~(3D-GF) from~\cite{Garcia2010}, and the proposed three-dimensional frequency selective reconstruction~(3D-FSR). In Figure~\ref{fig:sampling-patterns}, sections of all sampling patterns can be seen that have been used during the simulations. On the one hand, a static non-regular sampling pattern where the sampling positions do not change over time is applied and on the other hand, the proposed dynamic non-regular sampling pattern is applied where the sampling points are varying over time. For a sampling density of $25\%$, there are $\binom{4}{1} = 4$ combinations of complementary sampling patterns possible which are chosen in a way as shown in Figure~\ref{fig:sampling-patterns}. The same holds for a sampling density of $75\%$ where also $\binom{4}{3} = 4$ solutions exist. For a sampling density of $50\%$, however, $\binom{4}{2} = 6$ different complementary patterns can be realized which again are chosen in the order as seen in Figure~\ref{fig:sampling-patterns}. The combinations of these complementary sampling patterns are then repeated over time for all frames.

For the second approach to obtain both a high spatial and a high temporal resolution, two FRUC techniques are tested. Therefore, some frames of the original sequences are left out and all originally acquired frames have the full spatial resolution. For a sampling density of $25\%$ for example, only the first frame, the fifth frame, the ninth frame, and so on are available. In order to obtain the full temporal resolution, the missing intermediate frames are calculated by motion compensated linear averaging (MCLA)~\cite{Haan2010} and motion compensated three-dimensional frequency selective extrapolation (MC-3D-FSE)~\cite{Baetz2017}. 

The third approach to obtain both a high spatial and a high temporal resolution uses a sensor which gives a high temporal resolution at a low spatial resolution. In order to simulate such a camera sensor, for each frame of the original sequences always a group of $2\times2$ pixels is combined to one pixel. Four SR techniques are then applied to obtain the high spatial resolution. Iteratively Re-Weighted Minimization (IRWSR)~\cite{Koehler2016} and Fast and Robust Multi-Frame Super-Resolution (L1BTV)~\cite{Farsiu2014} are two multi-image SR techniques and Very Deep Super-Resolution (VDSR)~\cite{Kim2016} and Naive Bayes Super-Resolution Forest (NBSRF)~\cite{Salvador2015} are single-image SR techniques.

After the reconstruction of all sequences which have been acquired by these different approaches, the reconstruction quality is evaluated by comparing the reconstructed sequences to the original sequences using the peak-signal-to-noise ratio (PSNR).
The PSNR is not computed frame by frame, since sequences reconstructed by FRUC contain original frames which results in infinite PSNR values. Instead, the mean squared error is first calculated for the whole sequence followed by the calculation of the PSNR.
Additionally, a border of $14$ pixels in both spatial and temporal dimension gets excluded from calculation in order to avoid the influence of artifacts that some of the tested approaches introduce at the borders.

\begin{table}
	\caption{Parameters used by 3D-FSE and 3D-FSR.}
	\label{tab:fsr_parameter}
	\setlength{\tabcolsep}{4mm}
	\begin{tabularx}{\columnwidth}{Xc}
		\toprule
		Cube size                                         &    $4 \times 4 \times 4$     \\
		Border width                                      &             $14$             \\
		FFT size                                          &   $32 \times 32 \times 32$   \\
		Number of iterations per cube                     &            $500$             \\
		Decay factor $\hat{\rho}$                         &            $0.7$             \\
		Orthogonality deficiency compensation $\gamma$    &            $0.5$             \\
		Weighting of already reconstructed areas $\delta$ &            $0.5$             \\
		Adaptive frequency prior $\tau$ (only 3D-FSR)     &             $16$             \\
		Processing order for 3D-FSE                       & Optimized~\cite{Seiler2016b} \\
		Processing order for 3D-FSR                       &        Density-based         \\ \bottomrule
	\end{tabularx}
\end{table}

The first simulations show the performance of the proposed 3D-FSR. For the adaptive frequency prior which is a novel feature in 3D-FSR, however, a proper value for the control parameter $\tau$ has to be found first. In doing so, the training sequences from Figure~\ref{fig:test-set}, where a static non-regular sampling with different sampling densities of $25\%$, $50\%$, and $75\%$ is applied, are reconstructed by 3D-FSR using the parameters from Table~\ref{tab:fsr_parameter}. Then, different values for $\tau$ ranging from $\tau=0.1$ to $\tau=32$ are tested.
\begin{figure}
	\centering
%
%
\definecolor{mycolor1}{rgb}{0.20810,0.16630,0.52920}%
\definecolor{mycolor2}{rgb}{0.00596,0.40861,0.88284}%
\definecolor{mycolor3}{rgb}{0.06406,0.55699,0.82396}%
\definecolor{mycolor4}{rgb}{0.05897,0.68376,0.72539}%
\definecolor{mycolor5}{rgb}{0.39526,0.74590,0.52444}%
\definecolor{mycolor6}{rgb}{0.75249,0.73840,0.37681}%
\definecolor{mycolor7}{rgb}{0.99904,0.76531,0.21641}%
\definecolor{mycolor8}{rgb}{0.97630,0.98310,0.05380}%
\begin{tikzpicture}

\begin{axis}[%
width=0.828\columnwidth,
height=0.568\columnwidth,
at={(0\columnwidth,0\columnwidth)},
scale only axis,
log origin=infty,
xmin=0.5,
xmax=3.5,
xlabel={Sampling density},
xtick={1,2,3},
xticklabels={$25\%$,$50\%$,$75\%$},
ymin=20,
ymax=30,
ylabel={Average PSNR [dB]},
ymajorgrids,
axis background/.style={fill=white},
legend style={at={(0.5,1.25)},anchor=north,legend columns=4,legend cell align=left,align=left,draw=white!15!white}
]
\addplot[ybar,bar width=0.08,bar shift=-0.35,draw=black,fill=mycolor1,area legend] plot table[row sep=crcr] {%
1	20.3767865441898\\
2	22.7383824807592\\
3	26.960726255306\\
};
\addlegendentry{$\tau=0.1$};

\addplot[ybar,bar width=0.08,bar shift=-0.25,draw=black,fill=mycolor2,area legend] plot table[row sep=crcr] {%
1	23.0952643479838\\
2	26.2050083832476\\
3	29.3150841540142\\
};
\addlegendentry{$\tau=0.5$};

\addplot[ybar,bar width=0.08,bar shift=-0.15,draw=black,fill=mycolor3,area legend] plot table[row sep=crcr] {%
1	24.3115650290191\\
2	26.8778595186129\\
3	29.5625354996404\\
};
\addlegendentry{$\tau=1$};

\addplot[ybar,bar width=0.08,bar shift=-0.05,draw=black,fill=mycolor4,area legend] plot table[row sep=crcr] {%
1	24.8929197139011\\
2	27.1769091744824\\
3	29.6747157831166\\
};
\addlegendentry{$\tau=2$};

\addplot[ybar,bar width=0.08,bar shift=0.05,draw=black,fill=mycolor5,area legend] plot table[row sep=crcr] {%
1	25.1231063337231\\
2	27.308161220045\\
3	29.7284527348157\\
};
\addlegendentry{$\tau=4$};

\addplot[ybar,bar width=0.08,bar shift=0.15,draw=black,fill=mycolor6,area legend] plot table[row sep=crcr] {%
1	25.2041605445929\\
2	27.3636267704662\\
3	29.7565168574782\\
};
\addlegendentry{$\tau=8$};

\addplot[ybar,bar width=0.08,bar shift=0.25,draw=black,fill=mycolor7,area legend] plot table[row sep=crcr] {%
1	25.223328831268\\
2	27.3911343752059\\
3	29.7672627946194\\
};
\addlegendentry{$\tau=16$};

\addplot[ybar,bar width=0.08,bar shift=0.35,draw=black,fill=mycolor8,area legend] plot table[row sep=crcr] {%
1	25.2131562554856\\
2	27.4025295028847\\
3	29.7714191569866\\
};
\addlegendentry{$\tau=32$};

\end{axis}
\end{tikzpicture}%
	\caption{Evaluation of the parameter $\tau$ that is used for controlling the adaptive suppression of high-frequency basis functions depending on the number of available pixels in the reconstruction area $\mathcal{L}$. Average PSNR values are shown for various values of $\tau$, for the training set, and for different sampling densities.}
	\label{fig:evaluating-tau}
\end{figure}
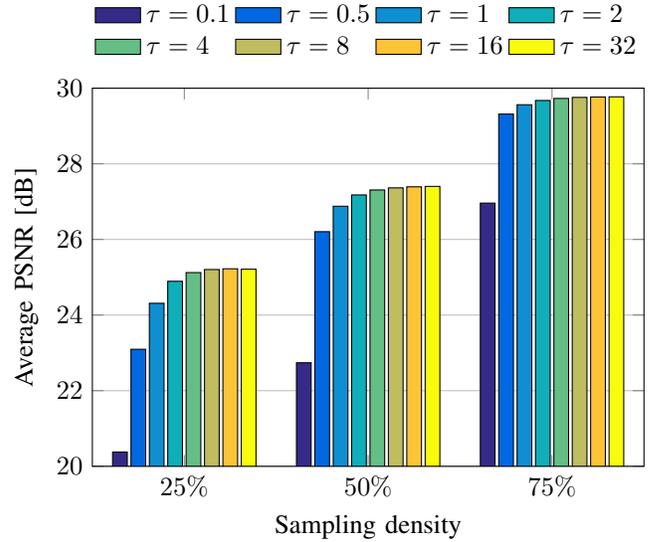
The results can be seen in Figure~\ref{fig:evaluating-tau} where the average PSNR is calculated over all training sequences and shown for different sampling densities. It can be seen that for all sampling densities, small values of $\tau$ lead to a poor reconstruction quality whereas for higher values of $\tau$ starting with $\tau = 16$ the reconstruction quality converges to its maximally reachable value. That is to say that $\tau=16$ is a good choice to achieve a reasonable reconstruction quality.

After choosing proper parameters for the proposed 3D-FSR, the performance of the 3D-FSR compared to the existing 3D-FSE~\cite{Seiler2016b} is evaluated. In doing so, the test sequences from Figure~\ref{fig:test-set} consisting of the HEVC ClassC and ClassD sequences, where again a static non-regular sampling with sampling densities of $25\%$, $50\%$, and $75\%$ is applied, are reconstructed by both 3D-FSR and 3D-FSE using the corresponding parameters from Table~\ref{tab:fsr_parameter}.
\begin{figure}
	\centering
%
%
\definecolor{mycolor2}{rgb}{0.40,0.80,0.20}%
\definecolor{mycolor1}{rgb}{1.00,0.80,0.00}%
\begin{tikzpicture}

\begin{axis}[%
width=0.85\columnwidth,
height=0.543\columnwidth,
at={(0\columnwidth,0\columnwidth)},
scale only axis,
log origin=infty,
xmin=0.5,
xmax=3.5,
xtick={1,2,3},
xticklabels={$25\%$,$50\%$,$75\%$},
xlabel={Sampling density},
ymin=28,
ymax=40,
ylabel={Average PSNR [dB]},
ymajorgrids,
axis background/.style={fill=white},
legend style={at={(0.03,0.97)},anchor=north west,legend cell align=left,align=left,draw=white!15!black}
]
\addplot[ybar,bar width=0.229,bar shift=-0.143,draw=black,fill=mycolor1,area legend] plot table[row sep=crcr] {%
1	29.410291088795\\
2	34.329022967413\\
3	39.0540652587767\\
};
\addlegendentry{3D-FSE~\cite{Seiler2016b}};

\addplot[ybar,bar width=0.229,bar shift=0.143,draw=black,fill=mycolor2,area legend] plot table[row sep=crcr] {%
1	29.9314180469059\\
2	34.4281053247623\\
3	39.0629517103023\\
};
\addlegendentry{proposed 3D-FSR};

\end{axis}
\end{tikzpicture}%
	\caption{Evaluation of the performance of the proposed 3D-FSR compared to the existing 3D-FSE~\cite{Seiler2016b} over the test sequences and for different sampling densities.}
	\label{fig:evaluating-3dfse-3dfsr}
\end{figure}
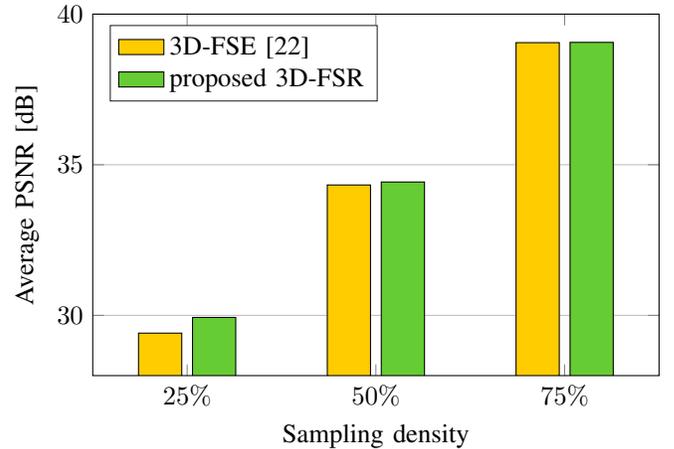
The results are shown in Figure~\ref{fig:evaluating-3dfse-3dfsr} where the average PSNR over all test sequences is calculated and shown for different sampling densities. It can be seen that 3D-FSR performs better than 3D-FSE for all sampling densities. For higher sampling densities, however, only marginal gains of up to $0.01$~dB can be reached. The highest gain with $0.51$~dB is achieved at small sampling densities. Therefore, it can be stated that 3D-FSR is especially suited for video data which is non-regularly sampled at low sampling densities.
3D-FSR is not only able to reconstruct non-regularly sampled video sequences with a high quality, it can also ensure the temporal consistency, since it uses three-dimensional Fourier basis functions. In doing so, smooth transitions in temporal direction can be guaranteed, as information from succeeding and preceding pixels are employed for the reconstruction.

Now, the proposed dynamic non-regular sampling concept is compared to a static non-regular sampling. Therefore, on the one hand, for each test sequence and each sampling density, a static non-regular sampling is applied, followed by a reconstruction by 3D-FSR. On the other hand, the proposed dynamic sampling is applied and also reconstructed by 3D-FSR.
\begin{table*}
	\caption{PSNR results in dB for the comparison of the proposed dynamic non-regular sampling to a static non-regular sampling using the proposed 3D-FSR for reconstruction.}
	\label{tab:psnr-results}
	\setlength{\tabcolsep}{4mm}
	\begin{tabularx}{\textwidth}{X|ccc|ccc|ccc}
		\toprule
		                \multirow{2}{*}{Sequences\vphantom{\Large X}}                  & \multicolumn{3}{c|}{\textbf{Sampling Density $\mathbf{25\%}$}} & \multicolumn{3}{c|}{\textbf{Sampling Density $\mathbf{50\%}$}} & \multicolumn{3}{c}{\textbf{Sampling Density $\mathbf{75\%}$}}                    \vspace{0.0cm} \\
		                                  & static\vphantom{\Large X} & dynamic &           $\Delta$           &  static  & dynamic &                   $\Delta$                    &  static  & dynamic &                                    $\Delta$                                    \\ \midrule
		\textbf{ClassC}                   &                          &         &                              &         &         &                                               &         &         &                                                                                \\
		BasketballDrill                   &         $32.02$          & $36.57$ &           $+4.55$            & $36.74$ & $40.86$ &                    $+4.12$                    & $41.63$ & $44.41$ &                                    $+2.87$                                     \\
		BQMall                            &         $29.98$          & $34.05$ &           $+4.07$            & $34.30$ & $38.66$ &               $\mathbf{+4.36}$                & $39.01$ & $42.24$ &                                $\mathbf{+3.23}$                                \\
		PartyScene                        &         $27.07$          & $31.63$ &       $\mathbf{+4.56}$       & $32.04$ & $35.59$ &                    $+3.55$                    & $36.90$ & $39.30$ &                                    $+2.40$                                     \\
		RaceHorses                        &         $30.45$          & $30.40$ &           $-0.05$            & $33.73$ & $33.64$ &                    $-0.09$                    & $37.43$ & $37.38$ &                                    $-0.05$                                     \\
		\addlinespace
		  \textbf{ClassD} &                          &         &                              &         &         &                                               &         &         &                                                                                \\
		BasketballPass                    &         $33.38$          & $36.53$ &           $+3.15$            & $37.86$ & $40.92$ &                    $+3.06$                    & $42.36$ & $44.75$ &                                    $+2.39$                                     \\
		BlowingBubbles                    &         $31.07$          & $33.43$ &           $+2.36$            & $35.00$ & $37.41$ &                    $+2.41$                    & $39.55$ & $40.98$ &                                    $+1.43$                                     \\
		BQSquare                          &         $24.96$          & $33.51$ &       $\mathbf{+8.55}$       & $31.25$ & $37.79$ &               $\mathbf{+6.54}$                & $37.05$ & $41.47$ &                                $\mathbf{+4.42}$                                \\
		RaceHorses                        &         $30.31$          & $30.34$ &           $+0.03$            & $33.61$ & $33.63$ &                    $+0.02$                    & $37.29$ & $37.35$ &                                    $+0.06$                                     \\ \midrule
		Average gain &                     ---     &   ---      &  $+3.40$                            &    ---     &   ---      &  $+3.00$                                             &    ---     &  ---       &    $+2.01$                                                                            \\\bottomrule
	\end{tabularx}	
\end{table*}
The detailed results are depicted in Table~\ref{tab:psnr-results} where for each test sequence the PSNR values are shown for the reconstruction after a static sampling and after the proposed dynamic sampling. It can be seen that for almost all sequences remarkable gains can be achieved when applying the proposed dynamic sampling. It is also evident that the gains decrease with a higher sampling density. The highest gain of $8.55$~dB is achieved for the reconstruction of the sequence {\itshape BQSquare} using a sampling density of $25\%$. This is a mostly static sequence with only few motion which is advantageous when using the proposed dynamic sampling. For static areas, a static sampling always misses the same position whereas the dynamic sampling acquires each position at least once within the range of four frames. On the other hand, when dealing with sequences with a lot of motion like {\itshape RaceHorses}, only marginal gains or even marginal losses are achieved. This is due to the reconstruction error when dealing with different alternating sampling patterns. That is to say that for such sequences the application of the proposed dynamic sampling almost yields the same results as a static sampling.
\begin{figure}
	\centering
%
%
\definecolor{mycolor1}{rgb}{0.3,0.6,0.9}
\definecolor{mycolor2}{rgb}{1.00,0.47,0.30}%
\begin{tikzpicture}

\begin{axis}[%
width=0.85\columnwidth,
height=0.481\columnwidth,
at={(0\columnwidth,0\columnwidth)},
scale only axis,
log origin=infty,
xmin=0.5,
xmax=7.5,
xtick={1,3,5,7},
xticklabels={2D-FSR~\cite{Seiler2015},3D-LIN~\cite{Watson1994},3D-GF~\cite{Garcia2010},3D-FSR},
xticklabel style={rotate=0},
ymin=28.5,
ymax=38,
ylabel={Average PSNR [dB]},
axis background/.style={fill=white},
ymajorgrids,
legend style={at={(0.03,0.97)},anchor=north west,legend columns=1,legend cell align=left,align=left,draw=white!15!black}
]
\addplot[ybar,bar width=0.229,bar shift=-0.143,fill=mycolor1,draw=black,area legend] plot table[row sep=crcr] {%
1	32.9703103520601\\
3	30.3984536541676\\
5	30.7249069628269\\
7	34.3733308509682\\
};
\addlegendentry{static sampling~\cite{Schoeberl2011}};

\addplot[ybar,bar width=0.229,bar shift=0.143,fill=mycolor2,draw=black,area legend] plot table[row sep=crcr] {%
1	32.987210785495\\
3	29.3813190070283\\
5	32.2930090436891\\
7	37.2010993068798\\
};
\addlegendentry{proposed dynamic sampling};

\end{axis}
\end{tikzpicture}%
	\caption{Evaluation of the performance of the proposed dynamic non-regular sampling compared to a static sampling. Averaged results over all test sequences and the three sampling densities $25\%$, $50\%$, and $75\%$ for each reconstruction algorithm.}
	\label{fig:evaluating-fix-var}
\end{figure}
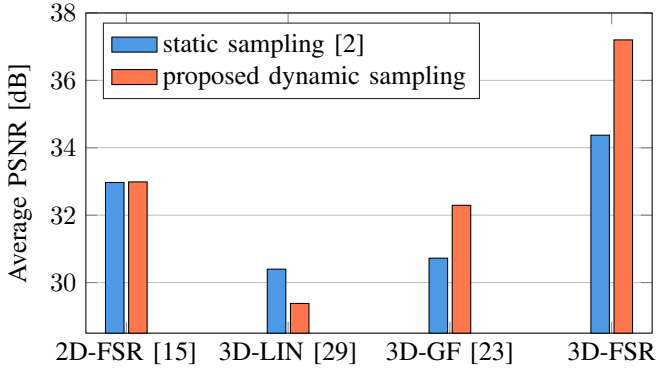
In Figure~\ref{fig:evaluating-fix-var}, the overall results for the comparison of a static non-regular sampling to the proposed dynamic non-regular sampling can be seen where the PSNR values are averaged for all test sequences and all sampling densities. Additionally, three other reconstruction algorithms for the sampled sequences are evaluated. The two-dimensional frequency selective reconstruction (2D-FSR) from~\cite{Seiler2015} performs a frame-wise reconstruction. The three-dimensional linear interpolation (3D-LIN) uses the built-in function \texttt{griddata} from MATLAB and is based on~\cite{Watson1994}. The three-dimensional gap filling (3D-GF) from~\cite{Garcia2010} is based on a penalized least squares method. In~\cite{Seiler2015}, it has been shown for the reconstruction of non-regularly two-dimensional sampled data that 2D-FSR performs better than other state-of-the-art reconstruction algorithms such as the statistically driven Steering Kernel Regression~\cite{Takeda2007} or the sparsity-based Wavelet Inpainting~\cite{Starck2010}. Therefore, only 2D-FSR is taken into account in this scenario. It can be clearly seen from Figure~\ref{fig:evaluating-fix-var} that algorithms like the 2D-FSR that do not exploit any temporal correlation cannot benefit from the dynamic sampling. On the other hand, other algorithms that exploit the third dimension for reconstruction like 3D-GF or the proposed 3D-FSR yield a better reconstruction quality when the proposed dynamic sampling is applied. In this case, an average gain in PSNR of $2.80$~dB can be achieved for the reconstruction by 3D-FSR. 
For the reconstruction by 3D-LIN a dynamic sampling is detrimental. This is due to the functionality of 3D-LIN. For a static sampling, 3D-LIN gets reduced to a frame-wise reconstruction. For the proposed dynamic sampling, 3D-LIN also gets reduced to a two-dimensional reconstruction, however, in temporal direction. 3D-LIN produces strong artifacts for areas with motion, since it performs no motion compensation. For a static background, however, 3D-LIN gives a good reconstruction quality, but overall, 3D-LIN is not suitable for the reconstruction of dynamic non-regularly sampled video data. 

\begin{figure*}
	\centering
	\def\svgwidth{\textwidth}
	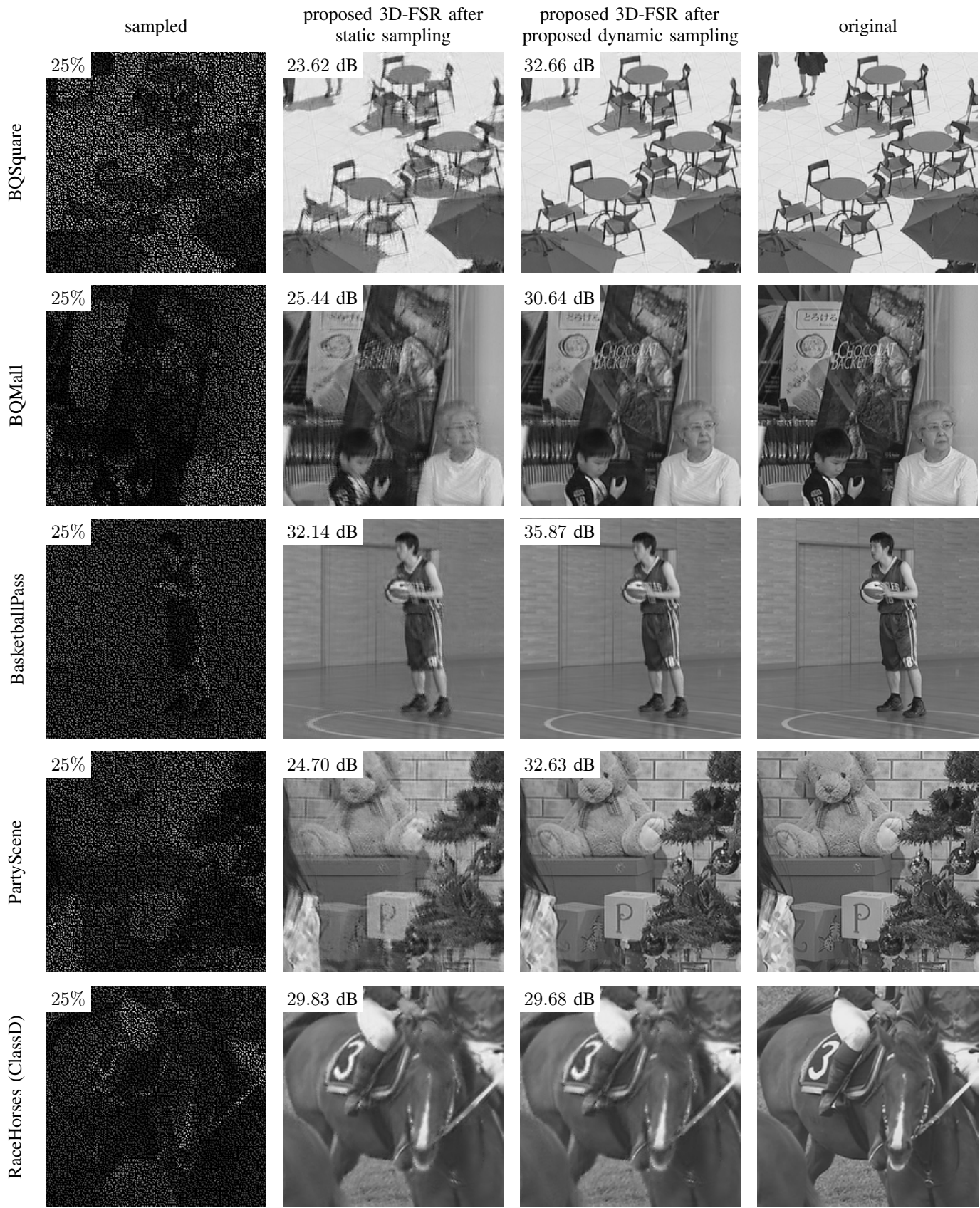
	\caption{Visual results for details of different test videos from the HEVC video data base showing a non-regular sampling density of 25\% and a reconstruction by 3D-FSR after both a static sampling and the proposed dynamic sampling. For each sequence, frame 25 is shown and the PSNR value is computed on the particular section of the frame. (Please pay attention that additional aliasing may be caused by printing or scaling. Best to be viewed enlarged on a monitor.)}
	\label{fig:image-examples}
\end{figure*}
\begin{table*}
	\caption{PSNR results in dB for the comparison of the proposed dynamic non-regular sampling with a subsequent reconstruction by 3D-FSR to other methods from the categories of FRUC and SR at a sampling density of $25\%$.}
	\label{tab:psnr-results-fruc-sr}
	\setlength{\tabcolsep}{3.5mm}
	\begin{tabularx}{\textwidth}{X|cc|cccc|c}
		\toprule
		\multirow{2}{*}{Sequences\vphantom{\Large X}} &   \multicolumn{2}{c|}{\textbf{Frame Rate Up-Conversion (FRUC)}}    &              \multicolumn{4}{c|}{\textbf{Super-Resolution (SR)}}               & \multicolumn{1}{c}{\textbf{Proposed}}                    \vspace{0.0cm} \\
		                                              & MCLA~\cite{Haan2010}\vphantom{\Large X} & MC-3D-FSE~\cite{Baetz2017} & IRWSR~\cite{Koehler2016} & L1BTV~\cite{Farsiu2014} & VDSR~\cite{Kim2016} & NBSRF~\cite{Salvador2015} &                                 3D-FSR                                  \\ \midrule
		\textbf{ClassC}                               &                                         &                          &                          &                         &                           &               &                                                          \\
		BasketballDrill                               &                 $26.00$                 &         $26.63$          &         $28.51$          &         $29.39$        & $29.65$ &          $30.45$          &                            $\mathbf{36.57}$                             \\
		BQMall                                        &                 $29.18$                 &         $30.41$          &         $24.60$          &         $25.05$        & $24.87$ &          $25.36$          &                            $\mathbf{34.05}$                             \\
		PartyScene                                    &                 $27.20$                 &         $28.11$          &         $21.80$          &         $22.42$        & $22.05$ &          $22.61$          &                            $\mathbf{31.63}$                             \\
		RaceHorses                                    &                 $21.90$                 &         $22.86$          &         $29.07$          &         $29.27$        & $28.02$ &          $28.76$          &                            $\mathbf{30.40}$                             \\
		\addlinespace
		  \textbf{ClassD}             &                                         &                          &                          &                         &                           &                                        &                                 \\
		BasketballPass                                &                 $35.39$                 &         $36.24$          &         $27.90$          &         $29.52$        & $28.16$ &          $28.88$          &                            $\mathbf{36.53}$                             \\
		BlowingBubbles                                &                 $26.69$                 &         $26.94$          &         $25.57$          &         $26.99$        & $26.41$ &          $27.11$          &                            $\mathbf{33.43}$                             \\
		BQSquare                                      &                 $32.20$                 &     $\mathbf{35.11}$     &         $20.70$          &         $21.12$        & $21.13$ &          $21.38$          &                                 $33.51$                                 \\
		RaceHorses                                    &                 $22.82$                 &         $23.78$          &         $27.83$          &         $29.36$        & $28.27$ &          $29.32$          &                            $\mathbf{30.34}$                             \\ \midrule
		Average PSNR                                  &                 $27.67$                 &         $28.76$          &         $25.75$          &         $26.64$        & $26.07$ &          $26.73$          &                            $\mathbf{33.31}$                             \\ \bottomrule
	\end{tabularx}	
\end{table*}
\begin{figure*}
	\centering
	\def\svgwidth{0.9\textwidth}
	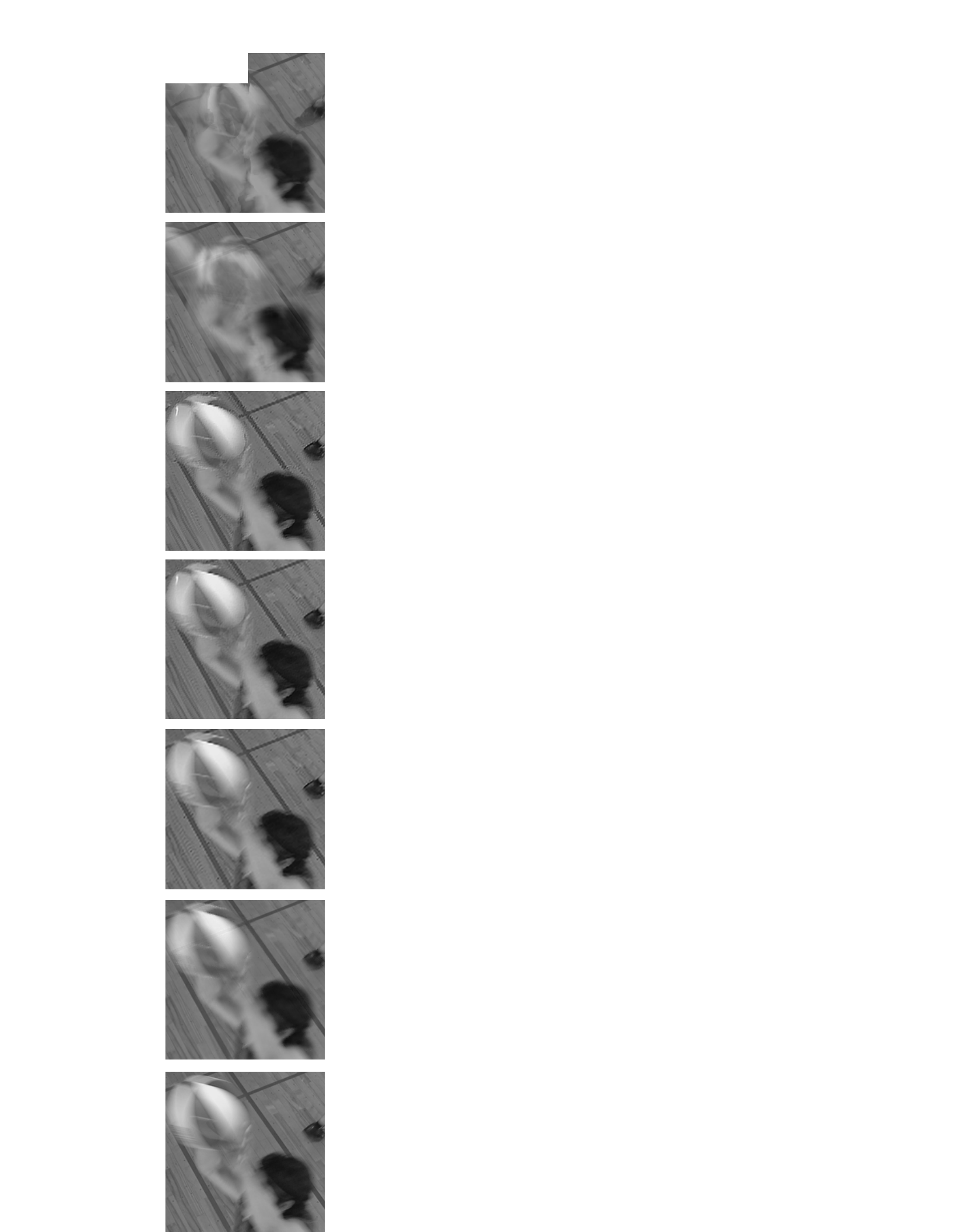
	\caption{Visual results for details of two test sequences from the HEVC video data base shown for different FRUC and SR techniques compared to the proposed dynamic sampling and for a sampling density of $25\%$. For each sequence, frame 27 is shown and the PSNR value is computed on the particular section of the frame. (Please pay attention that additional aliasing may be caused by printing or scaling. Best to be viewed enlarged on a monitor.)}
	\label{fig:image-examples-fruc-sr-dynamic}
\end{figure*}
The remarkable gains from Table~\ref{tab:psnr-results} and Figure~\ref{fig:evaluating-fix-var} are also visually noticeable and illustrated in Figure~\ref{fig:image-examples}. There, details of five sequences are shown where a sampling density of $25\%$ for both the proposed dynamic non-regular sampling and the static sampling is applied. After the reconstruction by 3D-FSR, the PSNR value is computed for the particular region of the shown frames. It can be clearly seen that by using the proposed dynamic sampling with the proposed 3D-FSR, the visual quality is significantly higher for almost every sequence. This is especially true for sequences with very few motion like {\itshape BQSquare} where the reconstruction of the tables and chairs almost reaches the quality of the original frame. Also other fine structures like the text in {\itshape BQMall} or the details in the other sequences can be reconstructed with a very high quality. The only case where the reconstruction quality is not higher is when the regarded sequence contains a lot of motion as for example in both {\itshape RaceHorses}. The reconstruction after the proposed dynamic sampling of such sequences, however, always is in the range of the reconstruction after a static sampling. These results show that by using the proposed dynamic non-regular sampling with a subsequent reconstruction by the also proposed 3D-FSR, sequences with both a high temporal and a high spatial resolution can be reconstructed. 

For a more comprehensive comparison, the proposed framework is now compared to other techniques from image processing that are also able to reconstruct such sequences. Therefore, the two FRUC techniques MCLA~\cite{Haan2010} and MC-3D-FSE~\cite{Baetz2017} and also the four SR techniques IRWSR~\cite{Koehler2016}, L1BTV~\cite{Farsiu2014}, VDSR~\cite{Kim2016}, and NBSRF~\cite{Salvador2015} are applied on the test sequences and compared to the proposed framework of dynamic non-regular sampling with a subsequent reconstruction by 3D-FSR. A sampling density of $25\%$ has been chosen in order to guarantee a fair comparison, since all approaches have to reconstruct the same amount of missing samples. They are, however, differently distributed in spatial and temporal direction. For FRUC, three out of four frames have to be estimated. For SR, the reconstruction of the high resolution frames is conducted from frames which are subsampled by a factor of two.
The results are shown in Table~\ref{tab:psnr-results-fruc-sr} where the PSNR values are given for each tested sequence and for the different techniques FRUC, SR, and the proposed dynamic non-regular sampling. It can be seen that for almost all sequences, using the proposed dynamic sampling yields a better quality compared to the other techniques. There are, however, scenarios where FRUC achieves a higher quality. This is the case for MC-3D-FSE~\cite{Baetz2017} and when the sequence contains almost no motion as for example in {\itshape BQSquare}. Since MC-3D-FSE performs a motion estimation, it does not have to interpolate missing samples when there is no motion, it can simply project these samples from adjacent frames. Nevertheless, compared to the best FRUC technique, the proposed dynamic non-regular sampling achieves an average gain of $4.55$~dB. Compared to the best SR technique, even average gains of up to $6.58$~dB are possible.
In Figure~\ref{fig:image-examples-fruc-sr-dynamic}, visual results are illustrated which also show that the proposed dynamic sampling yields a significantly higher quality than other methods from the categories of FRUC or SR. Only when there are mostly static regions in the sequence, FRUC gives a slightly higher quality. In general, FRUC methods like MCLA or MC-3D-FSE achieve a very high quality for these regions, however, when there are dynamic regions, the quality significantly decreases due to the error-prone motion estimation that is required for the projection of pixels in missing frames. With SR techniques like IRWSR, L1BTV, VDSR, or NBSRF, a high quality can be obtained for dynamic regions, however, for static regions only a poor quality can be reconstructed as most SR algorithms exploit pixel variations in order to super-resolve a video sequence. Using the proposed dynamic non-regular sampling, a very high quality can be reconstructed throughout the whole sequence independent of the motion.

\section{Conclusion \& Outlook}
\label{sec:conclusion}
In this paper, a novel strategy for modifying an image sensor in order to acquire both a high spatial and a high temporal resolution is proposed. The high temporal resolution is achieved by reading out only a subset of pixels per frame which allows for a higher throughput and therefore more frames per second. Subsequently, the so acquired incomplete frames are reconstructed by the also proposed three-dimensional frequency selective reconstruction (3D-FSR) in order to obtain the full high resolution sequence. The main contribution of this paper is the dynamic non-regular sampling which allows for each frame a dynamic readout of the sampling points. By using complementary sampling patterns, the reconstruction quality of 3D-FSR and other three-dimensional reconstruction algorithms can be significantly increased. That way, PSNR gains of up to $8.55$~dB can be achieved compared to only using a static sampling. Compared to other state-of-the-art techniques such as frame rate up-conversion or super-resolution which are also able to obtain sequences with both a high spatial and a high temporal resolution, an average gain in PSNR of up to $6.58$~dB is reached using the proposed dynamic non-regular sampling.

Future research aims at including motion estimation within this framework of dynamic non-regular sampling followed by a subsequent reconstruction. As shown in~\cite{Seiler2010a}, the reconstruction quality of the three-dimensional frequency selective extrapolation (3D-FSE) can be enhanced by adapting the model generation to the content of the sequence to be reconstructed. In doing so, motion estimation can be utilized to project pixels from adjacent frames instead of reconstructing missing pixels by sophisticated algorithms. Another possibility to adapt the model generation of the frequency selective reconstruction to the content of the data to be reconstructed has been shown in \cite{Jonscher2016a,Genser2017} where structural information has been used as a criterion. This information can also be used to assign different sampling densities to differently structured regions in the sequence.


%
%

\section*{Acknowledgment}
The authors gratefully acknowledge that this work has been supported by the Deutsche Forschungsgemeinschaft (DFG) under contract number KA 926/5-3.

\ifCLASSOPTIONcaptionsoff
  \newpage
\fi

\vspace{-0.5cm}
\begin{IEEEbiography}[{\includegraphics[width=1in,height=1.25in,clip,keepaspectratio]{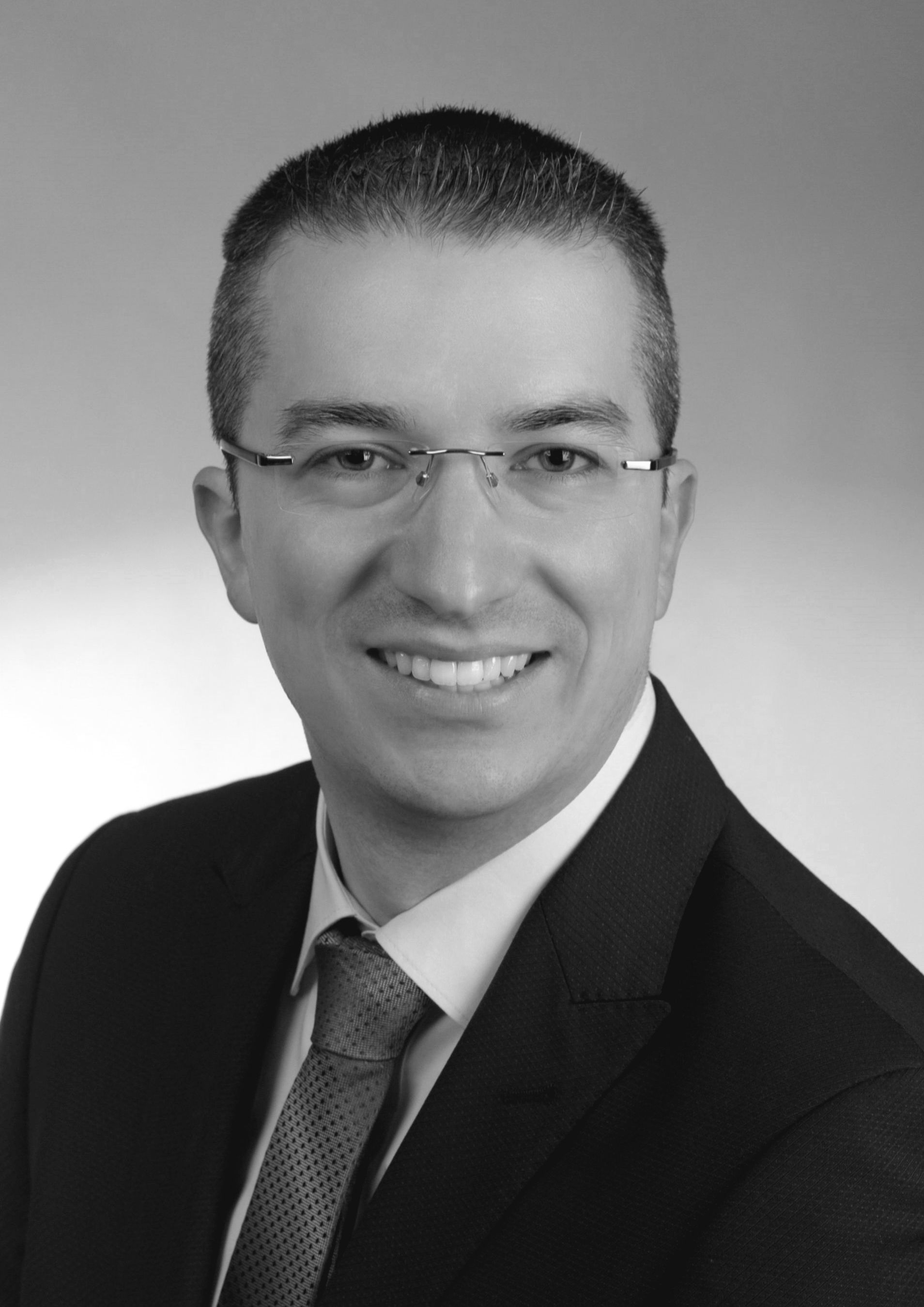}}]{Markus Jonscher}
	received the M.Sc. (Hons.) degree in systems of information and multimedia technology from Friedrich–Alexander–Universität	Erlangen-Nürnberg (FAU), Germany, in 2012. He then joined the Chair of Multimedia Communications and Signal Processing, where he conducted his research on image and video signal processing and reconstruction. He received a Top 10\% Best Paper Award at the IEEE International Conference on Image Processing (ICIP) in 2014.
	
	In 2018, he received the Dr.-Ing. degree in electrical engineering for his work on the reconstruction of non-regularly sampled image and video data. He is now the managing director of the Department for Electrical, Electronics and Communication Engineering (EEI), Erlangen, Germany.	
\end{IEEEbiography}
\vspace{-0.5cm}
\begin{IEEEbiography}[{\includegraphics[width=1in,height=1.25in,clip,keepaspectratio]{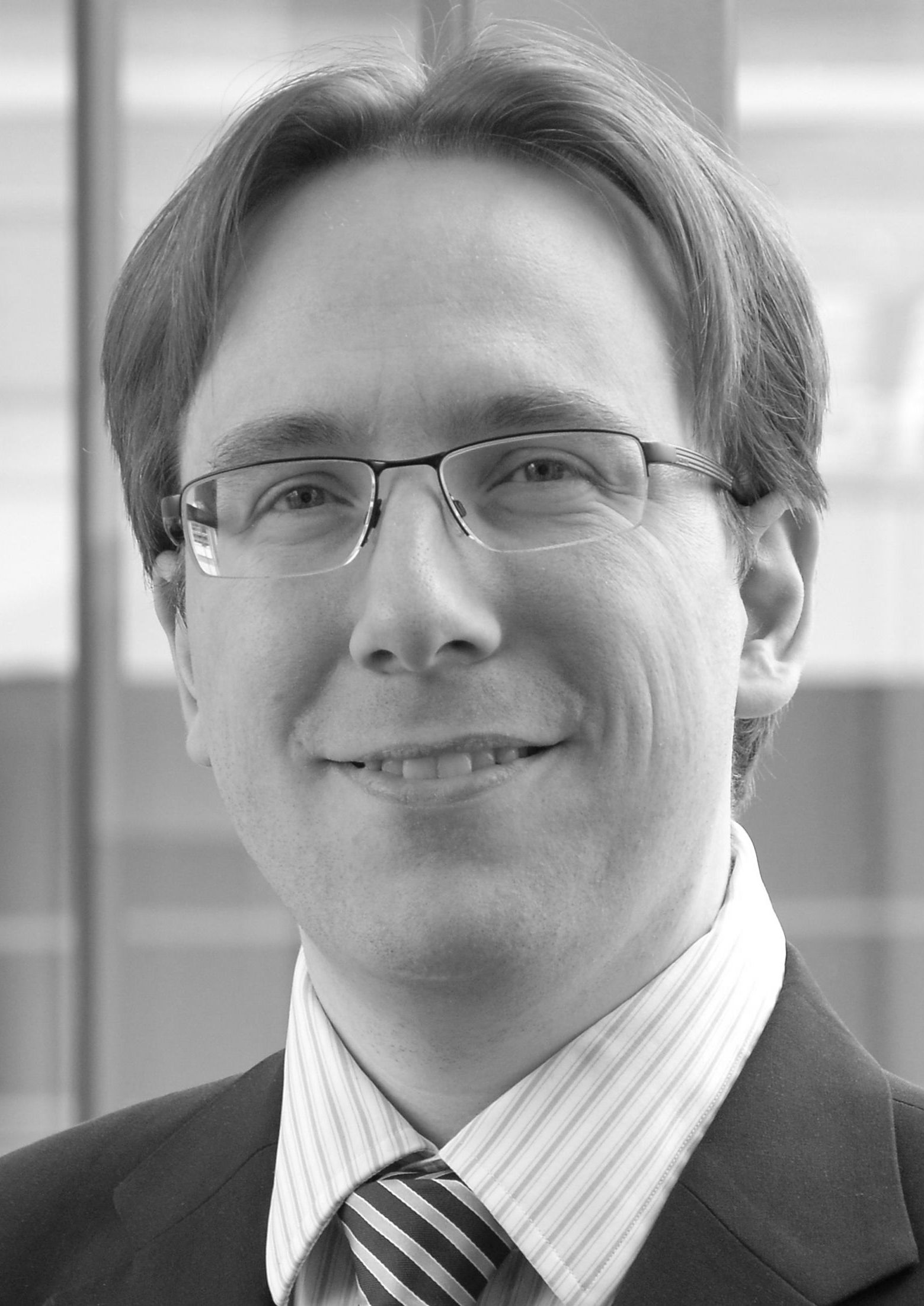}}]{Michael Schöberl}
	received a diploma in electrical engineering with honors in 2006 from the University of Erlangen-Nuremberg. He then joined the Fraunhofer IIS and started to work on high-quality imaging systems in the Electronic	Imaging department. In 2007, he joined the University of Erlangen-Nuremberg to continue the research on algorithms for	high-end camera systems in a close cooperation with Fraunhofer IIS. In 2013, he received a Ph.D. with honors in electrical engineering for his work on digital cameras.
	
	After working as the Chief Scientist in Electronic Imaging, he is now Head of the Group Imaging Solutions at the Fraunhofer IIS with a focus developing and implementing novel imaging technologies.	
\end{IEEEbiography}
\vspace{-0.5cm}
\begin{IEEEbiography}[{\includegraphics[width=1in,height=1.25in,clip,keepaspectratio]{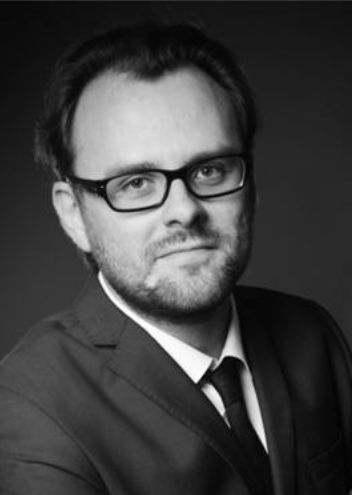}}]{Michel
		Bätz}
	received the Dipl.-Ing. degree in systems of information and multimedia technology and the Dr.-Ing. degree in electrical engineering from Friedrich-Alexander University Erlangen-N\"urnberg (FAU), Erlangen, Germany, in 2012 and 2018, respectively.	
	
	From 2012 to 2017, he was with the Chair of Multimedia Communications and Signal Processing at FAU, where he conducted his research on spatial, temporal, and radiometric resolution enhancement techniques for images and videos.
	Currently, he is with the Fraunhofer Institute for Integrated Circuits, Erlangen, Germany.
		
	His research interests include super-resolution as well as computational imaging with a particular focus on light field processing.
\end{IEEEbiography}
\begin{IEEEbiography}[{\includegraphics[width=1in,height=1.25in,clip,keepaspectratio]{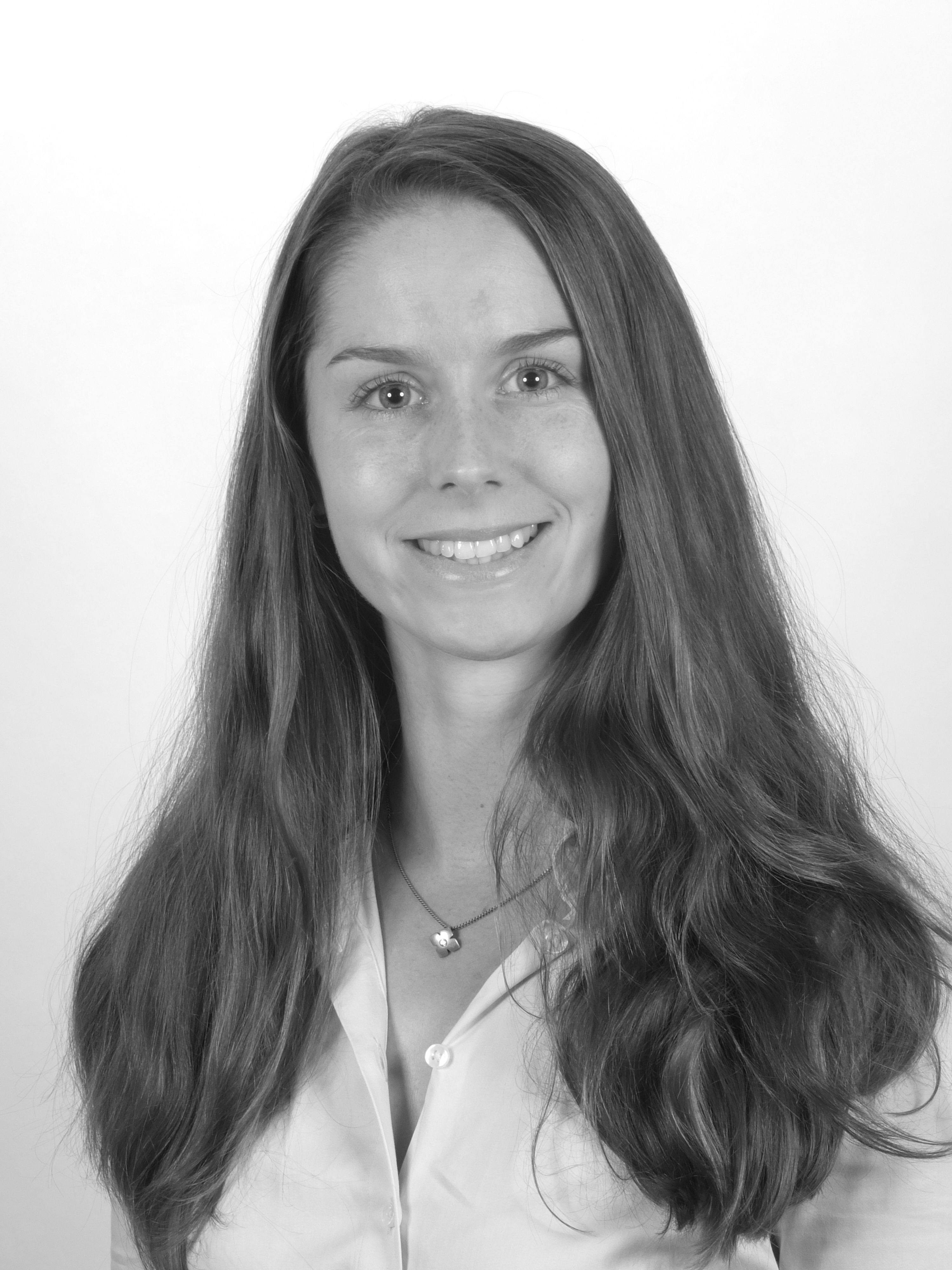}}]{Daniela Lanz}
	received the master’s degree in medical engineering from the Friedrich-Alexander-University Erlangen-Nürnberg (FAU), Erlangen, Germany, in 2015. Since 2015, she has been a Researcher with the Chair of Multimedia Communications and Signal Processing at FAU. Her current research interests include scalable lossless video coding, motion compensated wavelet lifting, graph-based signal processing, and related fields. She received a Teaching Award of the Faculty of Engineering in 2017.
\end{IEEEbiography}
\vspace{-0.5cm}
\begin{IEEEbiography}[{\includegraphics[width=1in,height=1.25in,clip,keepaspectratio]{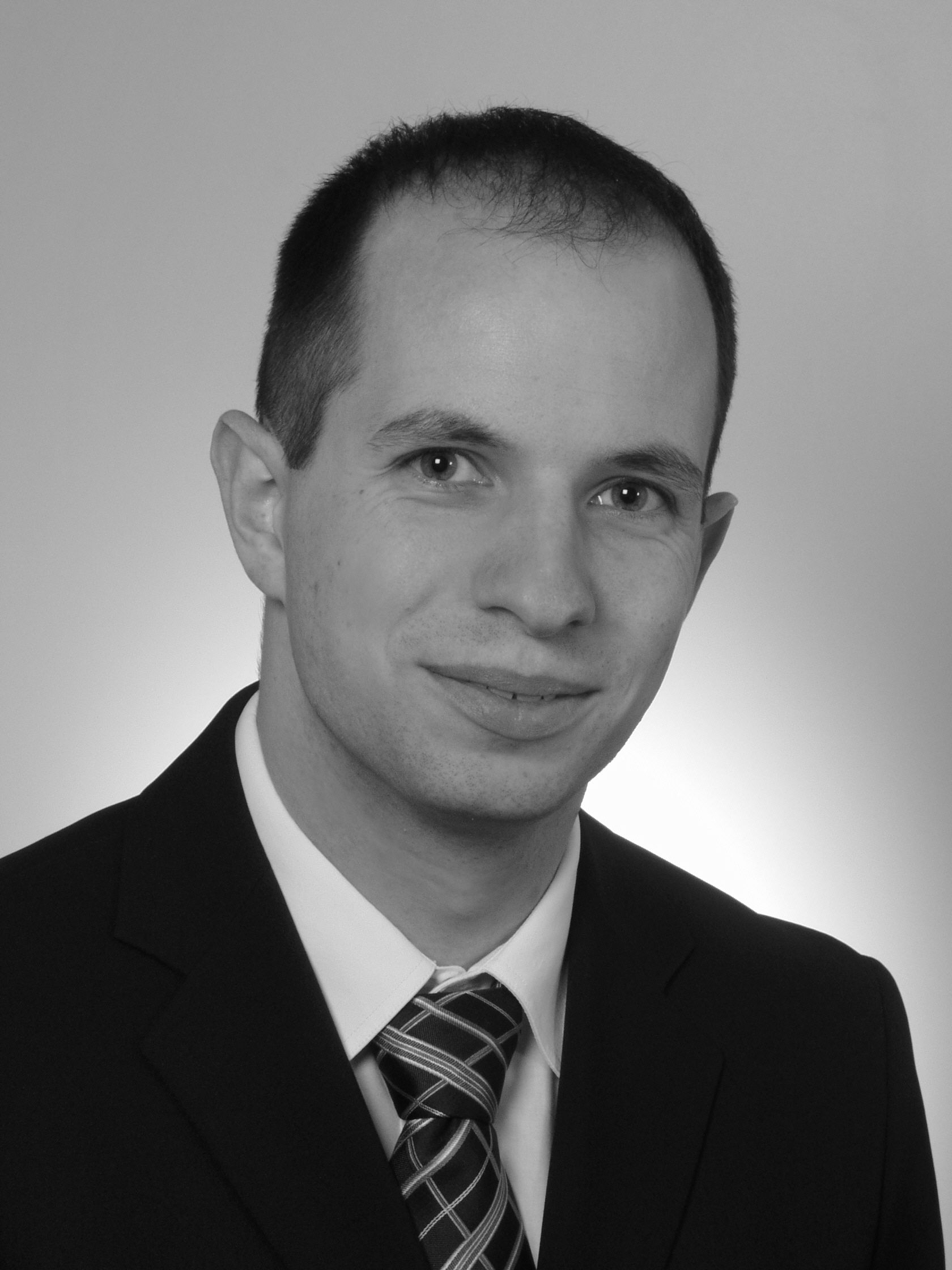}}]{Jürgen Seiler}
	(M'09-SM'16) is senior scientist and lecturer at the Chair of Multimedia Communications and Signal Processing at the Friedrich-Alexander Universit{\"a}t Erlangen-N{\"u}rnberg, Germany. There, he also received his habilitation degree in 2018, his doctoral degree in 2011, and his diploma degree in Electrical Engineering, Electronics and Information Technology in 2006.	
	
	He received the dissertation award of the Information Technology Society of the German Electrical Engineering Association as well as the dissertation award of the Staedtler-Foundation, both in 2012. In 2007, he received diploma awards from the Institute of Electrical Engineering, Electronics and Information Technology, Erlangen, as well as from the German Electrical Engineering Association. He also received scholarships from the German National Academic Foundation and the Lucent Technologies Foundation. He is the co-recipient of four best paper awards and he has authored or co-authored more than 80 technical publications.
		
	From 2012 to 2016, he was member of the Management Committee of EU COST Action IC1105 3D-ConTourNet - „3D Content Creation, Coding and Transmission over Future Media Networks“. For this Action, he also served as coordinator for Short Term Scientific Missions. In 2016, he received an LFUI-Guest Professorship at the University of Innsbruck.	
	His research interests include image and video signal processing, signal reconstruction and coding, signal transforms, and linear systems theory.
\end{IEEEbiography}
\vspace{-0.5cm}
\begin{IEEEbiography}[{\includegraphics[width=1in,height=1.25in,clip,keepaspectratio]{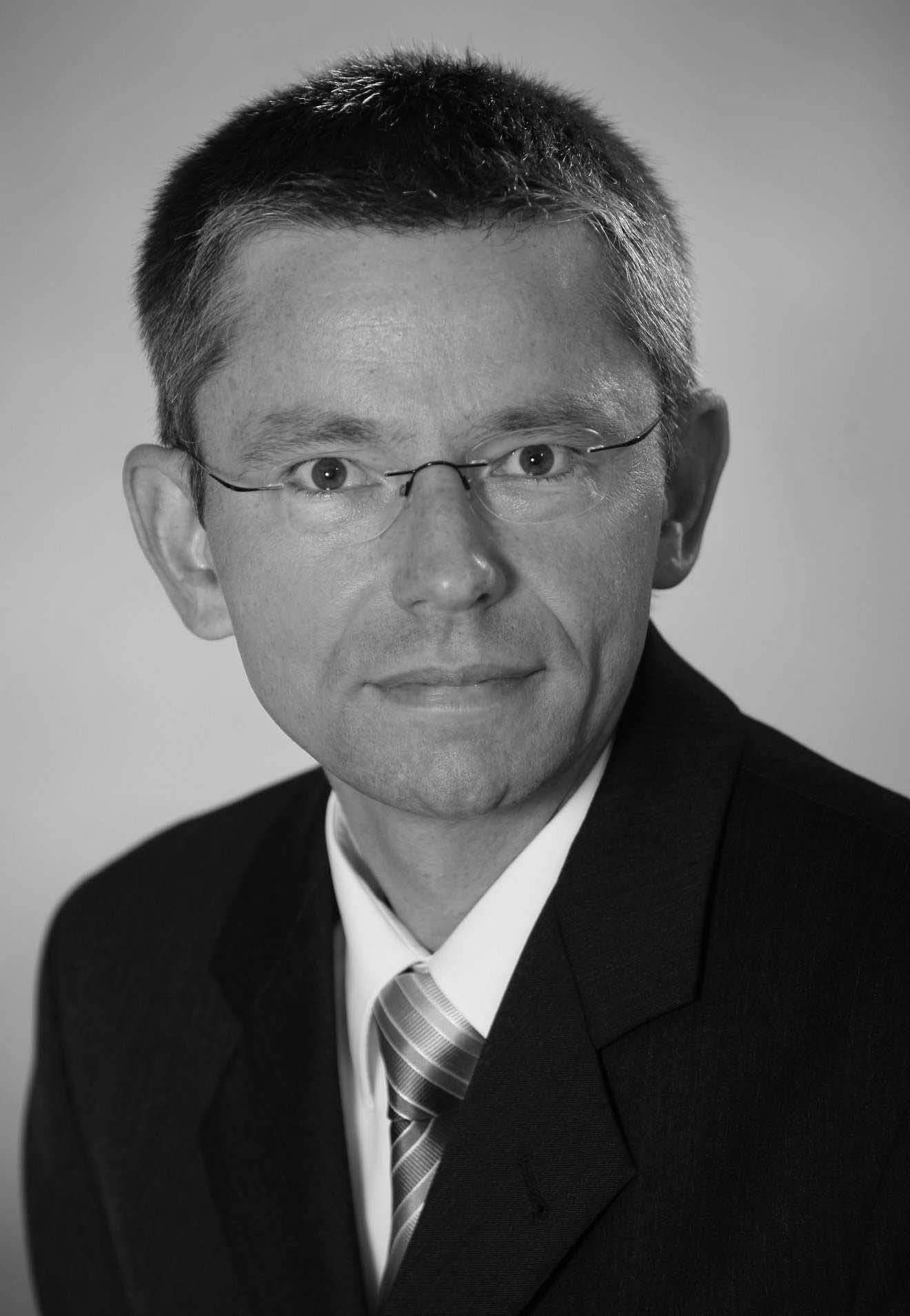}}]{André Kaup}
	(M’96–SM’99–F’13) received the Dipl.-Ing. and Dr.-Ing. degrees in electrical engineering from Rheinisch-Westfälische Technische Hochschule (RWTH) Aachen University, Aachen, Germany, in 1989 and 1995, respectively.
		
	He was with the Institute for Communication Engineering, RWTH Aachen University, from 1989 to 1995. He joined the Networks and Multimedia Communications Department, Siemens Corporate Technology, Munich, Germany, in 1995 and became Head of the Mobile Applications and Services Group in 1999. Since 2001 he has been a Full Professor and the Head of the Chair of Multimedia Communications and Signal Processing, University of Erlangen- Nuremberg, Erlangen, Germany. From 1997 to 2001 he was the Head of the German MPEG delegation. From 2005 to 2007 he was a Vice Speaker of the DFG Collaborative Research Center 603. From 2015 to 2017 he served as Head of the Department of Electrical Engineering and Vice Dean of the Faculty of Engineering. He has authored around 350 journal and conference papers and has over 100 patents granted or pending. His research interests include image and video signal processing and coding, and multimedia communication.	
	
	André Kaup is a member of the IEEE Multimedia Signal Processing Technical Committee, a member of the scientific advisory board of the German VDE/ITG, and a Fellow of the IEEE. He served as an Associate Editor for IEEE TRANSACTIONS ON CIRCUITS AND SYSTEMS FOR VIDEO TECHNOLOGY and was a Guest Editor for IEEE JOURNAL OF SELECTED TOPICS IN SIGNAL PROCESSING. From 1998 to 2001 he served as an Adjunct Professor with the Technical University of Munich, Munich. He was a Siemens Inventor of the Year 1998 and received the 1999 ITG Award. He has received several best paper awards, including the Paul Dan Cristea Special Award from the International Conference on Systems, Signals, and Image Processing in 2013. His group won the Grand Video Compression Challenge at the Picture Coding Symposium 2013 and he received the Teaching Award of the Faculty of Engineering in 2015. In 2018 he was elected full member of the Bavarian Academy of Sciences.
\end{IEEEbiography}
\end{document}